\documentclass[twocolumn,notrackchanges]{aastex62}
\pdfoutput=1 
\usepackage{amsmath,amstext}
\usepackage[T1]{fontenc}
\usepackage{apjfonts} 
\usepackage[figure,figure*]{hypcap}
\usepackage{cleveref}
\usepackage{float}
\usepackage{hyperref}
\usepackage[utf8]{inputenc}
\usepackage{newunicodechar} 
\usepackage{gensymb}

\newcommand\kms{\mathrm{km~s^{-1}}}

\newcommand\Myr{\mathrm{Myr}}

\newcommand\K{\mathrm{K}}
\newcommand\m{\mathrm{m}}
\newcommand\cm{\mathrm{cm}}
\newcommand\pc{\mathrm{pc}}
\newcommand\Tk{T_\mathrm{K}}
\newcommand\Tex{T_\mathrm{ex}}
\newcommand\NH{\mathrm{NH_3}}
\newcommand\dv{\sigma_\mathrm{v}}
\newcommand\vlsr{\mathrm{v_{LSR}}}

\newcommand\Mhz{\mathrm{MHz}}
\newcommand\kHz{\mathrm{kHz}}

\newunicodechar{°}{\degree}

\begin{document}

\shorttitle{Dense gas kinematics and a narrow filament in OMC1}
\shortauthors{Monsch et al. (2018)}

\title{Dense gas kinematics and a narrow filament in the Orion A OMC1 region using $\NH$}

\author{Kristina Monsch}
\affiliation{Universit\"ats-Sternwarte, Ludwig-Maximilians-Universit\"at M\"unchen, Scheinerstr.~1, 81679 M\"unchen, Germany}
\email{kristina.monsch@gmail.com}

\author{Jaime E. Pineda}
\affiliation{Max-Planck-Institut f\"ur extraterrestrische Physik, Giessenbachstrasse 1, 85748 Garching, Germany}

\author{Hauyu Baobab Liu}
\affiliation{European Southern Observatory (ESO), Karl-Schwarzschild-Str. 2, 85748 Garching, Germany}

\author{Catherine Zucker} 
\affiliation{Harvard-Smithsonian Center for Astrophysics, 60 Garden St., Cambridge, MA, USA}

\author{Hope How-Huan Chen} 
\affiliation{Harvard-Smithsonian Center for Astrophysics, 60 Garden St., Cambridge, MA, USA}

\author{Kate Pattle} 
\affiliation{Jeremiah Horrocks Institute, University of Central Lancashire, Preston, Lancashire PR1 2HE, UK}
\affiliation{Institute of Astronomy and Department of Physics, National Tsing Hua University, Hsinchu, Taiwan}

\author{Stella S. R. Offner}
\affiliation{Department of Astronomy, The University of Texas at Austin, 2515 Speedway, Stop C1400, Austin, TX 78712-1205, USA}

\author{James Di Francesco}
\affiliation{National Research Council of Canada, Herzberg Institute of Astrophysics, 5071 West Saanich Road, Victoria, BC,
V9E 2E7, Canada}
\affiliation{Department of Physics and Astronomy, University of Victoria, 3800 Finnerty Rd., Victoria, BC, V8P 5C2, Canada}

\author{Adam Ginsburg}
\affiliation{Jansky fellow of the National Radio Astronomy Observatory, 1003 Lopezville Rd, Socorro, NM 87801 USA}

\author{Barbara Ercolano} 
\affiliation{Universit\"ats-Sternwarte, Ludwig-Maximilians-Universit\"at M\"unchen, Scheinerstr.~1, 81679 M\"unchen, Germany}
\affiliation{Excellence Cluster Origin and Structure of the Universe, Boltzmannstr.2, 85748 Garching, Germany}

\author{H\'ector G. Arce}
\affiliation{Department of Astronomy, Yale University, P.O. Box 208101, New Haven, CT 06520-8101, USA}

\author{Rachel Friesen}
\affiliation{National Radio Astronomy Observatory, 520 Edgemont Rd, Charlottesville, VA 22903, USA}
\affiliation{Dunlap Institute for Astronomy \& Astrophysics, 50 St. George St., Toronto ON Canada M5S 3H4}

\author{Helen Kirk}
\affiliation{National Research Council of Canada, Herzberg Institute of Astrophysics, 5071 West Saanich Road, Victoria, BC,
V9E 2E7, Canada}
\affiliation{Department of Physics and Astronomy, University of Victoria, 3800 Finnerty Rd., Victoria, BC, V8P 5C2, Canada}

\author{Paola Caselli}
\affiliation{Max-Planck-Institut f\"ur extraterrestrische Physik, Giessenbachstrasse 1, 85748 Garching, Germany}

\author{Alyssa A. Goodman} 
\affiliation{Harvard-Smithsonian Center for Astrophysics, 60 Garden St., Cambridge, MA, USA}

\begin{abstract}
We present combined observations of the $\NH~(J,K)=(1,1)$ and $(2,2)$ inversion transitions toward OMC1 in Orion~A obtained by the \textit{Karl G. Jansky} Very Large Array (VLA) and the $100~\m$ \textit{Robert C. Byrd} Green Bank Telescope (GBT). With an angular resolution of $6''$ (0.01 pc), these observations reveal with unprecedented detail the complex filamentary structure extending north of the active Orion BN/KL region in a field covering $\sim 6'\times 7'$. 
We find a $0.012~\pc$ wide filament within OMC1, with an aspect ratio of $\sim$37:1, that was missed in previous studies. Its orientation is directly compared to the relative orientation of the magnetic field from the \textit{James Clerk Maxwell} Telescope BISTRO survey in Orion~A. We find a small deviation of $\sim11\degree$ between the mean orientation of the filament and the magnetic field, suggesting that they are almost parallel to one another. The filament's column density is estimated to be 2--3 orders of magnitude larger than the filaments studied with \textit{Herschel} and is possibly self-gravitating given the low values of turbulence found. 
We further produce maps of the gas kinematics by forward modeling the hyperfine structure of the $\NH~(J,K)=(1,1)$ and $(2,2)$ lines. The resulting distribution of velocity dispersions peaks at $\sim0.5~\kms$, close to the subsonic regime of the gas. This value is about $0.2~\kms$ smaller than previously measured in single-dish observations of the same region, suggesting that higher angular and spectral resolution observations will identify even lower velocity dispersions that might reach the subsonic turbulence regime in dense gas filaments.
\end{abstract}


\section{Introduction}
\label{sec:introduction}

Ammonia ($\NH$) is one of the most useful molecules available for studying the physical and chemical properties of cold and dense gas found in prestellar cores, where stars are likely to form. 
At the typically low temperatures ($\lesssim 10$\,--\,$20~\K$) and high hydrogen densities ($\gtrsim 10^4$\,--\,$10^5~\mathrm{cm^{-3}}$) found in these regions, severe depletion of CO and its isotopologues occurs due to freeze-out onto dust grains \citep{Caselli+1999, Bacmann+2002, Bergin+2002, Tafalla+2002, Tafalla+2004}. Under such conditions, the amount of nitrogen-bearing species, such as $\NH$ and $\mathrm{N_{2}H^+}$, increases \citep[e.g.,][]{BerginLanger1997, Aikawa+2001, diFrancesco+2007, Lippok+2013}. Due to their relatively high abundances in the interstellar medium (ISM) and their large number of transitions that are sensitive to a wide range of excitation conditions, $\NH$ and $\mathrm{N_{2}H^+}$ are very useful probes of dense molecular gas \citep{HoTownes1983}.
Moreover, the innermost regions of prestellar cores are well traced by these species \citep[e.g.,][]{diFrancesco+2007,Pety+2017}, making them a powerful tool to study the conditions found in star-forming regions. 

The population of the metastable $\NH~(J=K)$ levels are mainly excited by collisions, which makes ammonia an ideal molecular thermometer. Owing to its many hyperfine lines, kinetic information can be obtained with high accuracy from $\NH$ observations \citep{HoTownes1983, Tafalla+2002, Rosolowsky+2008, Friesen+2009, Lippok+2013, MangumShirley2015, FriesenPineda+2017}.

In this context, the nearby star-forming regions located in the Gould Belt \citep{Herschel1847, Gould1879} have been mapped at many wavelengths, identifying for example their young stellar population \citep[e.g.,][]{Preibisch+2005, Meggeath+2012, Meggeath+2016} or the structure of embedded filaments and dense cores \citep[e.g.,][]{Andre+2010, Arzoumanian+2011, Palmeirim+2013}. One of these regions is the Orion Molecular Cloud (OMC) complex, hosting a great variety of dark clouds, bright nebulae, and young stars. 
The OMC hosts two giant molecular clouds, Orion~A and Orion~B, which are connected by an extended, ridge-like structure of quiescent gas along the entire complex \citep{GenzelStutzki1989, Womack+1993}. 
Dense cores and filaments are embedded in these ridges, where active star formation is likely to take place. 
Located almost centrally within the Integral-Shaped Filament (ISF) of Orion~A is the Orion Nebula Cluster (ONC) \citep{O'dell2001, Menten+2007}. 
Parallax measurements toward the ONC (and a few detections toward the OMC) using Very Long Baseline Interferometry (VLBI) led to a distance of (388$\pm$5)~pc \citep{Kounkel+2017}. 
Although this value is slightly lower than previous measurements, with either fewer parallax measurements \citep{Menten+2007} or using a different technique \citep{Schlafly2014,Schlafly2015}, this is also the most recent determination for OMC1.
Certainly, future distance determinations using \textit{Gaia} will help to resolve this issue.

Within the OMC1 region, the highly active Becklin-Neugebauer and Kleinmann-Low region (hereafter Orion BN/KL) embedded in OMC1 is of particular interest \citep[e.g.,][]{GenzelStutzki1989, JohnstoneBally1999, Bally+2005, Crockett+2010, Neill+2013}. 
This part of Orion has been subject to extensive studies over the last few decades, 
either focusing on its characterization on large scales \citep[e.g.,][]{FriesenPineda+2017, Hacar+2017, Hacar+2018} or on the internal structure of Orion BN/KL and its local environment \citep[e.g.,][]{Tang+2010, Bally+2011, Bally+2017, Goddi+2011, Zapata+2011, Goicoechea+2015, Hirota+2015}.
There are, however, only a few studies that address the region surrounding Orion BN/KL on larger scales at high resolution, especially in high-density molecular tracers such as $\NH$. 

\citet{WisemanHo1996_nature, WisemanHo1998} were the first to study the $\NH$ emission in OMC1 at high angular ($8''$) resolution, using the Very Large Array (VLA). They found extended clumpy filaments, which they referred to as ``fingers'', in a region spanning $0.5~\pc$ toward the north and west of Orion BN/KL. These are separated into at least two major velocity components at $\sim 8~\kms$ and $10~\kms$, which appear to overlap in Orion BN/KL. However, spatially extended $\NH$ emission was filtered out in the VLA observations of \citet{WisemanHo1996_nature, WisemanHo1998}.
The absence of this component prohibits the precise measurement of physical quantities, such as the widths of filaments, spectral line ratios, or their corresponding line widths.
In particular, relatively faint structures are difficult to analyze correctly in the VLA-only map, due to the confusion from the negative intensity bowl resulting from negative side lobes surrounding nearby bright sources.
To amend these problems, we take advantage of Green Bank Telescope (GBT) data taken from the \textit{Green Bank Ammonia Survey} \citep[GAS,][]{FriesenPineda+2017}, which provide the complementary short-spacing information.
By reprocessing the archival VLA data with current mosaicking techniques and combining them with the single-dish GBT observations (see \S\,\ref{sec:imgcomb}), it is now possible to be more sensitive to both large- and small-scale structures in dense gas within OMC1.

In this paper, we present new results on the dense gas kinematics and structure of OMC1 in Orion~A using $\NH~(J,K)=(1,1)$ and $(2,2)$ observations.
Details on the data reprocessing are given in \S\,\ref{sec:observations}. In \S\,\ref{sec:methods}, we describe the analysis of the resulting data cube, as well as the spectral line fit used to model the $\NH~(J,K)=(1,1)$ and $(2,2)$ lines. The results and discussion are presented in \S\,\ref{sec:results} and \S\,\ref{sec:discussion}, respectively. 
The summary and conclusions follow in \S\,\ref{sec:summary}. \\

\section{Observations}
\label{sec:observations}

\begin{figure*}[h!]
\centering
\includegraphics[width=0.9\linewidth]{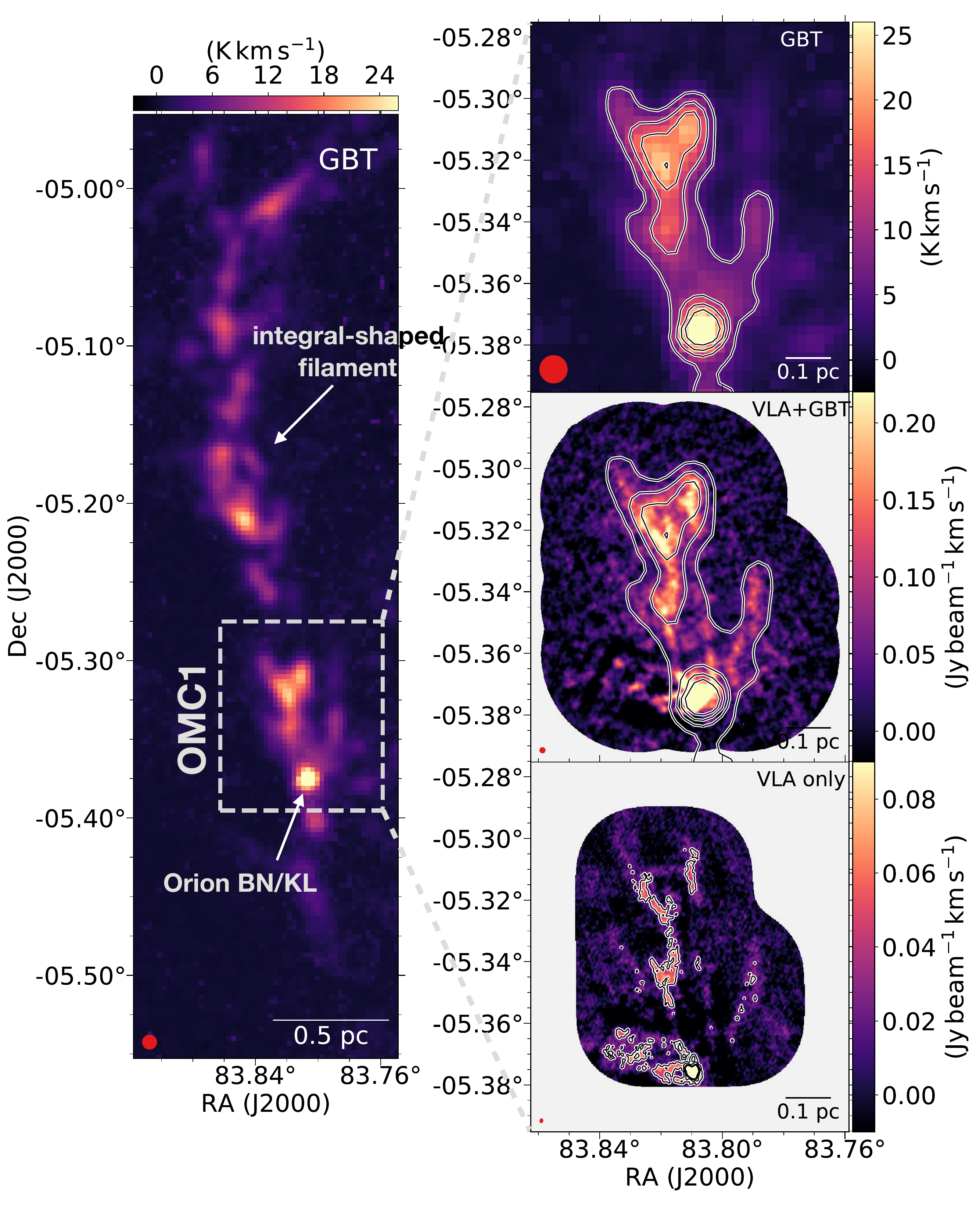}
\caption{Integrated intensity maps of the $\NH$ (1,1) line for Orion~A and OMC1. The corresponding beam sizes and scale bars are shown in the lower left and right corners, respectively. The color scales are in units of the integrated brightness temperature and flux density for the corresponding observation. \textit{Left:} Large-scale structure as observed by the GBT at $32''$ resolution, taken from the \textit{Green Bank Ammonia Survey} \citep[GAS,][]{FriesenPineda+2017}. The integrated intensity map was calculated within a velocity range that includes the emission of all hyperfine components in Orion~A. 
\textit{Top right:} Same observation as in the left panel, but zoomed into the dashed box, covering a region of $\sim6'\times7'$. 
\textit{Center right:} integrated $\NH~(1,1)$ intensity map of OMC1, calculated within $5.5$\,--\,$11.5\,\kms$ from the combined data cube at $6''$ resolution. 
\textit{Bottom right:} VLA-only $\NH~(1,1)$ integrated intensity map of the same region with an initial angular resolution of $\sim$4.5\arcsec$\times$3.5\arcsec \citep{WisemanHo1996_nature,WisemanHo1998}. It was calculated in a range spanning $4~\kms$ to include the emission of the main component only.
The contours show the extent of the GBT (upper two panels) and VLA (lower panel) $\NH~(1,1)$ integrated intensity at the $[5,10,15,20]-\sigma_{\mathrm{rms}}$ levels, where $\sigma_{\mathrm{rms}}$ is the rms-error estimated from emission-free pixels in the GBT- or VLA-only map, respectively.}
  \label{fig:full_gbt_vs_vla}
\end{figure*}

\subsection{Karl G. Jansky Very Large Array (VLA)}
\label{sec:VLA}

We used archival observations of $\NH$ $(J,K)=(1,1)$ and $(2,2)$ emission from OMC1 in the Orion~A molecular cloud, obtained using the VLA. 
These observations, as well as their extensive analysis, were already presented by \citet{WisemanHo1996_nature, WisemanHo1998}. Therefore, we refer the reader to these previous studies for a detailed description of the observational setup and the data reduction process, which we will only briefly summarize here. Both $\NH$ hyperfine inversion transitions were observed simultaneously in the VLA D-configuration, covering a field of view of about $4\arcmin\times8\arcmin$ at a final resolution of $8''$.
The total bandwidth of $3.125~\Mhz$ ($39~\kms$) covers the main hyperfine component as well as the inner pair of satellite components for each transition. The 127 channels are separated by $23.7~\kHz$, corresponding to a velocity resolution of $0.3~\kms$.
Flagging and calibration of the data were performed with \textsc{Casa} \citep{CASA}.

\subsection{Robert C. Byrd Green Bank Telescope (GBT)}
\label{sec:GBT}

The GBT data were observed as part of the \textit{Green Bank Ammonia Survey} (GAS) and presented in \cite{FriesenPineda+2017}.
In contrast to the VLA data, the observations obtained by the GBT have a nearly eight times wider bandwidth of 23.44~MHz, covering the main hyperfine component as well as the inner and outer pairs of satellite components. With a beam full width at half maximum (FWHM) of nearly 32\arcsec, the GBT is ideal for analyzing the large-scale structure of OMC1 at high spectral resolution ($5.7~\kHz$, corresponding to $\sim 0.07~\kms$). For more details on the observational setup and the corresponding data reduction, see \cite{FriesenPineda+2017}.

\subsection{Combination of the VLA and GBT observations}
\label{sec:imgcomb}

Empirically, it has been known that jointly deconvolving single-dish and interferometric images yields the sharpest combined images. However, joint deconvolution procedures \citep[e.g., the Maximum Entropy Method implemented in the {\tt mosmem} task of the \textsc{Miriad} software package, cf.][]{Miriad_software} in general do not preserve the overall fluxes measured with single-dish observations alone. This is because deconvolution algorithms are either not flux-conserving or still subject to edge effects. Another type of approach is to perform a linear weighted sum of the Fourier transformed single-dish and interferometric images \citep[e.g., using the {\tt immerge} task in \textsc{Miriad}, or the {\tt feather} task in \textsc{Casa}, cf.][]{CASA}. These approaches preserve the overall fluxes from single-dish observations. However, output images often suffer from the same defects as the interferometric images (e.g., strips, negative bowls), making the combined images more limited in terms of intensity dynamic range. The interferometric imaging defects are particularly serious in our case of imaging the ONC with the VLA, since structures in this field are crowded. In addition, the {\it uv} sampling was limited by the close to zero declination of this source, and the ``Y''-shaped array configuration. Either the non-preserved overall fluxes in the joint deconvolution approaches, or the imaging defects in the linear weighted-sum approaches do not act uniformly on individual spectral channels or spectral lines, and can therefore considerably bias our measurements of line widths and our derivation of gas temperatures and column densities from fitting line ratios. To generate images that are optimal for our scientific purposes by making use of the existing software packages, we tentatively adopted a hybrid approach.

The GBT images were regridded to match the VLA spectral resolution of $0.3~\kms$ and then deconvolved using the \texttt{clean} task in \textsc{Miriad} before converting into visibilities using the \texttt{uvrandom} and \texttt{uvmodel} tasks. 
The GBT and VLA visibilities were then jointly imaged using \texttt{invert} and \texttt{mosmem} within \textsc{Miriad} \citep[e.g.,][]{Vogel+1984}.
To ensure total flux conservation, we recombined the joint GBT and VLA image from the previous step with the GBT image using \texttt{immerge} to ensure that the total flux is recovered. 
This produced an image with a channel width of $0.3~\kms$ and a beam size of $\sim$4.5\arcsec$\times$3.5\arcsec, that was smoothed down to 6\arcsec\ to increase the sensitivity to line emission. We note that this resolution is higher than the $8''$ achieved by \citet{WisemanHo1996_nature,WisemanHo1998}. 
This is mainly a result of the significantly improved mosaic imaging and the combination with single dish data, that allow us to reach a higher signal-to-noise than the VLA-only data for the same beam. 
The mean rms-noise levels, $\sigma_{\mathrm{rms}}$, are $4.9~\mathrm{mJy~beam^{-1}}$ and $16.7~\mathrm{mJy~beam^{-1}}$ for the resulting $\NH~(1,1)$ and (2,2) channel maps, respectively.

Figure~\ref{fig:full_gbt_vs_vla} shows the $\NH~(1,1)$ integrated intensity maps of the single-dish observations (left and upper right panel) and the final data cube resulting from the combination process described above (center right panel). For comparison, we additionally show the recalibrated VLA-only data in the lower right panel with an initial angular resolution of $\sim4.5''\times3.5''$. The left panel corresponds to an extended view of the Orion~A ridge as observed in the GAS survey at 32\arcsec\ resolution. The dashed, white box highlights the 6\arcmin$\times$7\arcmin\ region covering OMC1 for which the combined data cube has been obtained. 
The right panels zoom into this region for three different data sets: the GBT-only map (top), the combined VLA+GBT map (center) and the VLA-only map (bottom). The direct comparison immediately shows the increase in image resolution both of small- and large-scale structures resulting from the combination of single-dish and interferometric data. 

\section{Analysis}
\label{sec:methods}

\begin{figure}[h!]
\centering
\includegraphics[width=\linewidth]{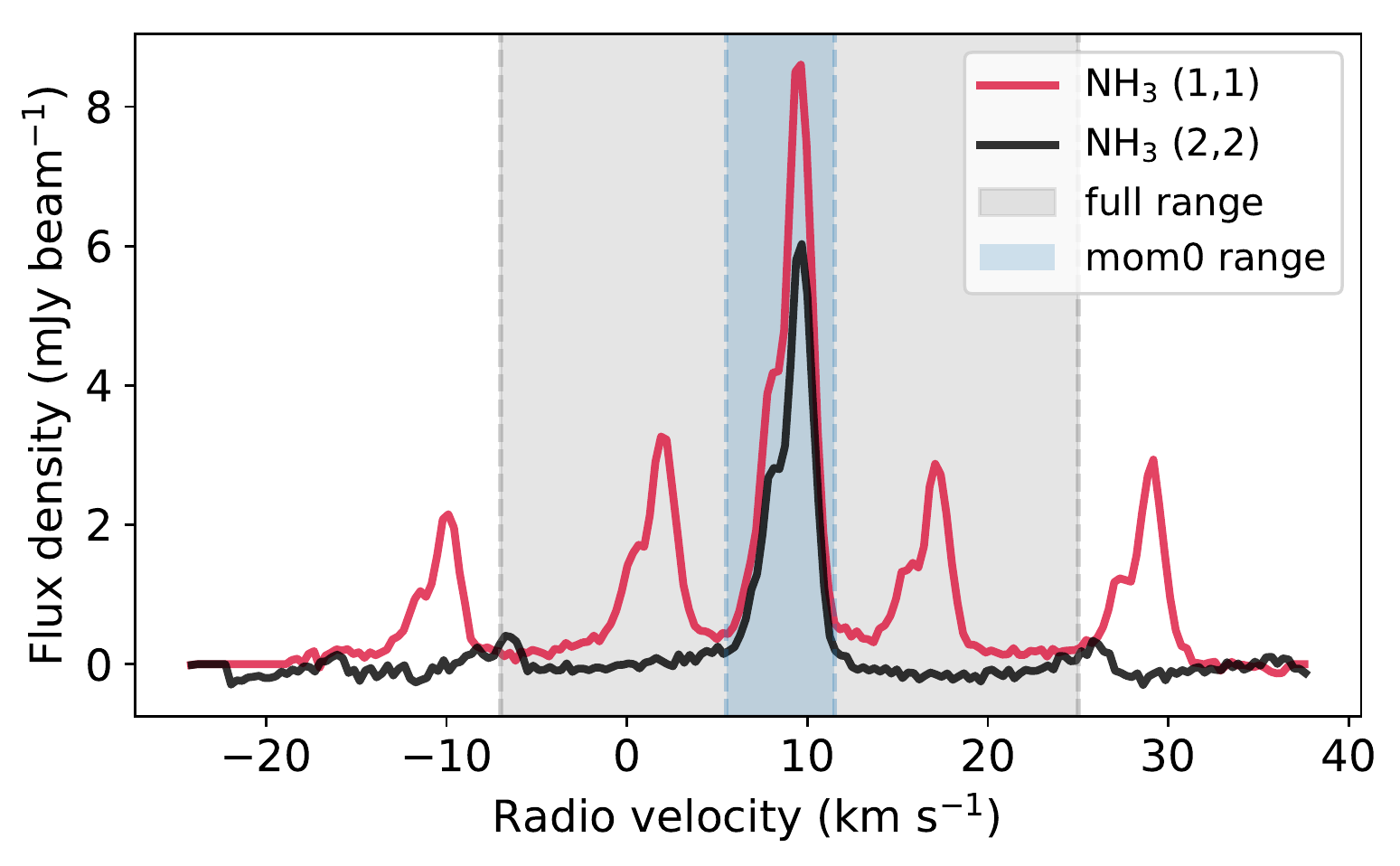}
\caption{Mean spectrum of the $\NH~(1,1)$ (red) and $(2,2)$ (black) lines from the combined VLA and GBT data. The gray shade highlights the spectral region between $-7.1~\kms$ and $25.1~\kms$ that was used for the analysis presented in this paper. The blue range shows the range between $5.5~\kms$ and $11.5~\kms$, in which the integrated intensity map was computed. This narrow range only includes the emission of the bright main component to highlight the overall gas distribution in OMC1. }
  \label{fig:mean_spectrum}
\end{figure}

Many steps that were applied in the following analysis have been discussed in detail in \citet{FriesenPineda+2017}, therefore we will only give a brief summary here. 
We analyzed the final data cubes of the $\NH\,(1,1)$ and $(2,2)$ hyperfine transitions simultaneously using \texttt{PySpecKit} \citep{pyspeckit}. 
The channels between $-7.1~\kms$ and $25.1~\kms$ were extracted from the combined data cube to remove the outer edges of the spectrum, including the second pair of satellite components. This step was necessary, as the combined spectra have higher rms-noise levels in that region. 
The resulting spectral data cube was then used to calculate the integrated intensity (or ``moment 0'') map for both $\NH$ transitions between $5.5~\kms$ and $11.5~\kms$. This range includes the emission of the bright main component only and highlights the overall gas structure in the region. In addition, we masked out all voxels\footnote{pixels in a three-dimensional data cube} outside of the 20\% primary beam response. 
Figure~\ref{fig:mean_spectrum} shows the mean spectrum of the combined $\NH~(1,1)$ and $\NH~(2,2)$ maps, as well the velocity ranges used for the here presented analysis.

The first moment map, a proxy for the centroid velocity $\vlsr$, was calculated only over channels where the line was detected with a signal-to-noise $\mathrm{SNR}>3$. Here, $\mathrm{SNR}=T_{\mathrm{peak}}/\sigma_\mathrm{rms}$, where $T_{\mathrm{peak}}$ is the peak brightness temperature and $\sigma_\mathrm{rms}$ is the rms-error in the $\NH\,(1,1)$ cube at a given pixel. This ensured that only pixels with a strong signal were included into the calculation, because the resulting map was then provided as an input guess to the line fit.
Close to the BN/KL region, the lines are wide and the continuum emission is strong enough to make the moment calculation uncertain. Unfortunately, the bandwidth is narrow and the spectral resolution is coarse, making the determination of a proper baseline impractical. Therefore, all pixels with too large mean continuum levels ($>3~\sigma_{\mathrm{rms}}$) were masked out.
For the remaining analysis, however, it was sufficient to keep all pixels with $\mathrm{SNR}>1$. 

Each of the $\NH\,(1,1)$ and $(2,2)$ spectra were then fitted simultaneously using the \verb+cold_ammonia+ model implemented in \texttt{PySpecKit}, which is based on previous work by \citet[]{Rosolowsky+2008, Friesen+2009, Pineda+2010, MangumShirley2015}. See also \citet{FriesenPineda+2017} for a detailed discussion of the spectral line fitting procedure itself. In this model, the emission in each pixel is assumed to arise from a homogeneous slab of uniform kinetic temperature $\Tk$, excitation temperature $\Tex$, $\NH$ column density $N(\NH)$, velocity dispersion $\dv$ and centroid velocity $\vlsr$ \citep{Swift+2005, Rosolowsky+2008, Pineda+2011}. 
The initial set of guesses used are $\Tk=27.1~\K$, $\Tex=8.5~\K$, $\log_{10}(N(\NH)/\cm^{-2})=14.1$, $\dv=0.6~\kms$ and $\vlsr=\texttt{mom1}$, where \texttt{mom1} corresponds to the first moment map of the $\NH\,(1,1)$ hyperfine transition. These values correspond to the median ones determined from the GAS data in OMC1. 
We note that instead of providing uniform initial guesses for the entire region, we could have used the parameter maps obtained from the GAS data line fit. However, due to different masking criteria used for the GAS results as well as high uncertainties in several GAS-pixels within OMC1 (close to BN/KL), this procedure results in too many pixels with poor fits, and therefore we settled for an intermediate approach which produced more stable fits. 
In addition, we used a single value for the velocity dispersion guess for the whole map to ensure that we could fit single components in regions with blended lines in the single-dish GAS data.
To account for the different rms-noise levels of both $\NH$ transitions, an error cube consisting of the respective $1\,\sigma_{\mathrm{rms}}$ errors for each transition was provided to the fit.
Finally, the observed line widths\footnote{Here we refer to the FWHM of a spectral line as the line width, defined as $\mathrm{FWHM} = \Delta \mathrm{v} = 2\sqrt{2 \ln 2}~\dv$.}, $\Delta \mathrm{v_{obs}}$, were corrected for artificial broadening resulting from the spectral resolution of $\Delta \mathrm{v_{res}}=0.3~\kms$ using the relation $\Delta \mathrm{v_{corr}}=\sqrt{\Delta \mathrm{v^2_{obs}}-\Delta \mathrm{v^2_{res}}}$ as discussed in \citet{Friesen+2009}.

\begin{deluxetable}{lccccc}[h]
\tablecolumns{6}
\tablewidth{0pt}
\tablecaption{Upper and lower limits and rms-errors ($\sigma_{\mathrm{fit}}$) used for the flagging of the data. The last line shows additional flagging criteria from the other parameters, as discussed in \S\,\ref{sec:methods}. \label{tab:flagging}}
\tablehead{\colhead{parameter} & \colhead{$\Tk$} & \colhead{$\Tex$} & \colhead{$N(\NH)$} & \colhead{$\dv$} & \colhead{$\vlsr$}\\
\colhead{unit} & \colhead{($\K$)} & \colhead{($\K$)} & \colhead{($\cm^{-2}$)} & \colhead{($\kms$)} & \colhead{($\kms$})}
\startdata
lower limit & 2.8 & 2.8 & $10^{12.0}$ & 0.04 & 5.5 \\
upper limit & 50.0 & 30.0 & $10^{17.0}$ & 5.0 & 11.0\\
$\sigma_{\mathrm{fit}}$ & 5.0 & 3.0 & $10^{0.5}$ & 0.1 & 0.1\\
add. $\sigma_{\mathrm{fit}}$ flag & $\vlsr$, $\dv$ & $\Tk$ & $\Tex$, $\Tk$ & $\vlsr$ & \nodata \\
\enddata
\end{deluxetable}

Not all pixels yield a reliable full set of fitted parameters. 
Only the parameters that are not well constrained are then removed, by comparing their corresponding associated uncertainty, $\sigma_{\mathrm{fit}}$, to the maximum allowed uncertainty listed in Table~\ref{tab:flagging}. 
This is possible, because some parameters are fairly independent. 
A reliable determination of $\vlsr$ and $\dv$ requires a good fit of the $\NH\,(1,1)$ line only, and if the centroid velocity ($\vlsr$) has an associated uncertainty larger than $0.1~\kms$ (a fraction of the channel width), then both $\vlsr$ and $\dv$ are removed from the results. 
$\Tex$, $\Tk$, and $N(\NH)$ further require a good fit of both $\NH\,(1,1)$ and $\NH\,(2,2)$ lines \citep{HoTownes1983}. 
An accurate determination of the kinetic temperature, $\Tk$, requires a good measurement of the ratio between the $\NH\,(1,1)$ and $\NH\,(2,2)$ lines, and therefore it should have a well determined $\vlsr$ and $\dv$. 
The excitation temperature, $T_{ex}$, is constrained by the ratio of the different hyperfine components in addition to the total line intensity of the $\NH\,(1,1)$ line. However, an inaccurate kinetic temperature yields an erroneous excitation temperature and therefore we require $\sigma_{\mathrm{fit}}<5~\K$ for $\Tk$.
An accurate determination of the column density of ammonia, $N(\NH)$, requires a good determination of both temperatures, kinetic and excitation \citep[see e.g.,][]{Friesen+2009,HoTownes1983}.
These and other flagging criteria, are summarized in Table~\ref{tab:flagging}.
Finally, any remaining isolated pixels or clumps unconnected to larger-scale features, most likely corresponding to noisy data points, are also removed.

The combined data cubes of the $\NH\,(1,1)$ and $(2,2)$ transitions as well as the resulting parameter maps are provided as FITS-files on Harvard dataverse (\url{https://doi.org/10.7910/DVN/QLD7TC}).

\begin{figure*}[]
\centering
\includegraphics[width=\linewidth]{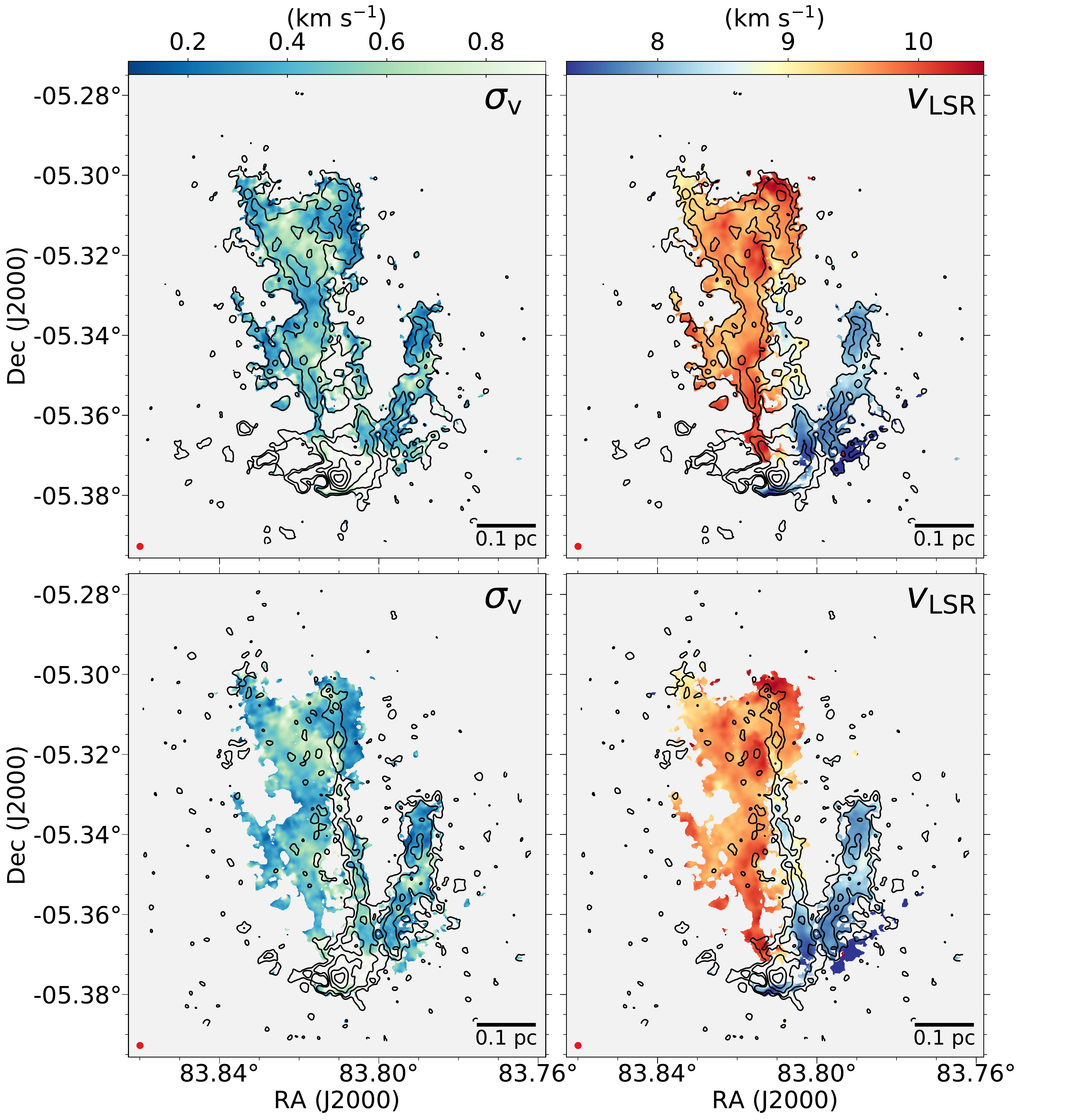}
\caption{Results of the spectral line fit, after pixels with uncertain fits were removed using the flagging conditions listed in Table \ref{tab:flagging}. \textit{Left:} velocity dispersion $\dv$, which was corrected for the channel width of $0.3~\kms$. \textit{Right:} centroid velocity $\vlsr$.
The contours in the upper two panels are drawn at the $[3, 6, 12,...]-\sigma_{\mathrm{rms}}$ levels of the $\NH\,(1,1)$ integrated intensity map of the final data cube. The lower two panels show contours at the same levels for the mean $\sigma_{\mathrm{rms}}$ within the narrow spectral range between $8.1$--$8.4~\kms$, to highlight the appearance of the fainter filaments.}
  \label{fig:fit_vel}
\end{figure*}

\section{Results}
\label{sec:results}

\subsection{Kinematics}
\label{sec:kinematics}

The hyperfine components of both $\NH$ transitions were fitted simultaneously using the forward model approach discussed in \S\,\ref{sec:methods}. Figure~\ref{fig:fit_vel} shows the corrected velocity dispersions $\dv$ (left) and centroid velocities $\vlsr$ (right) in OMC1 with two different sets of contours overlaid. In the upper two panels, contours are drawn at the $[3, 6, 12,...]-\sigma_{\mathrm{rms}}$ levels of the $\NH\,(1,1)$ integrated intensity map, while the contours in the lower two panels are drawn at the same levels of the mean $\sigma_{\mathrm{rms}}$ within the narrow spectral range of $8.1$--$8.4~\kms$. This intends to highlight the appearance of the fainter filaments that will be discussed in more detail in \S\,\ref{sec:mom0}. 
The velocity dispersion ranges from $\sim0.1~\kms$ to $\sim0.8~\kms$ across the entire field, reaching a mean value of $(0.5\pm0.04)~\kms$. This is on average about $0.2~\kms$ smaller than those measured from single-dish observations of OMC1 alone \citep{FriesenPineda+2017}.

The velocity of the gas along the line of sight, $\vlsr$, in the right panel of Figure~\ref{fig:fit_vel} shows a striking bi-modality. The velocity ranges from $< 7.8~\kms$ in the southwestern part of OMC1 to $>10.2~\kms$ in the northern part. Both cloud components likely overlap in the Orion BN/KL region, which has been masked out due to its high continuum emission. In addition, a velocity gradient along the filamentary structure at $\alpha \sim 83.81\degree$ becomes apparent, highlighted by the contours in the lower two panels of Figure~\ref{fig:fit_vel}. It connects the southern and northern parts of OMC1 not only spatially but also in terms of velocity.
We finally note that the overall large velocity gradient found in OMC1 is part of an even larger-scale velocity gradient in the entire Orion~A region, in which OMC1 is almost centrally embedded. These velocities range from $13~\kms$ north of OMC1 to lower values of $7~\kms$ to $9~\kms$ in the southernmost part of Orion~A and have been observed both in $\NH$ and $\mathrm{N_2H^+}$ emission \citep{FriesenPineda+2017,Hacar+2017}. However, immediately south of Orion BN/KL lies a higher-velocity region ($10$\,--\,$11~\kms$), contributing to the `V-shaped' velocity gradient observed by \citet{Hacar+2017}. 

\begin{figure}[h!]
\centering
\includegraphics[width=\linewidth]{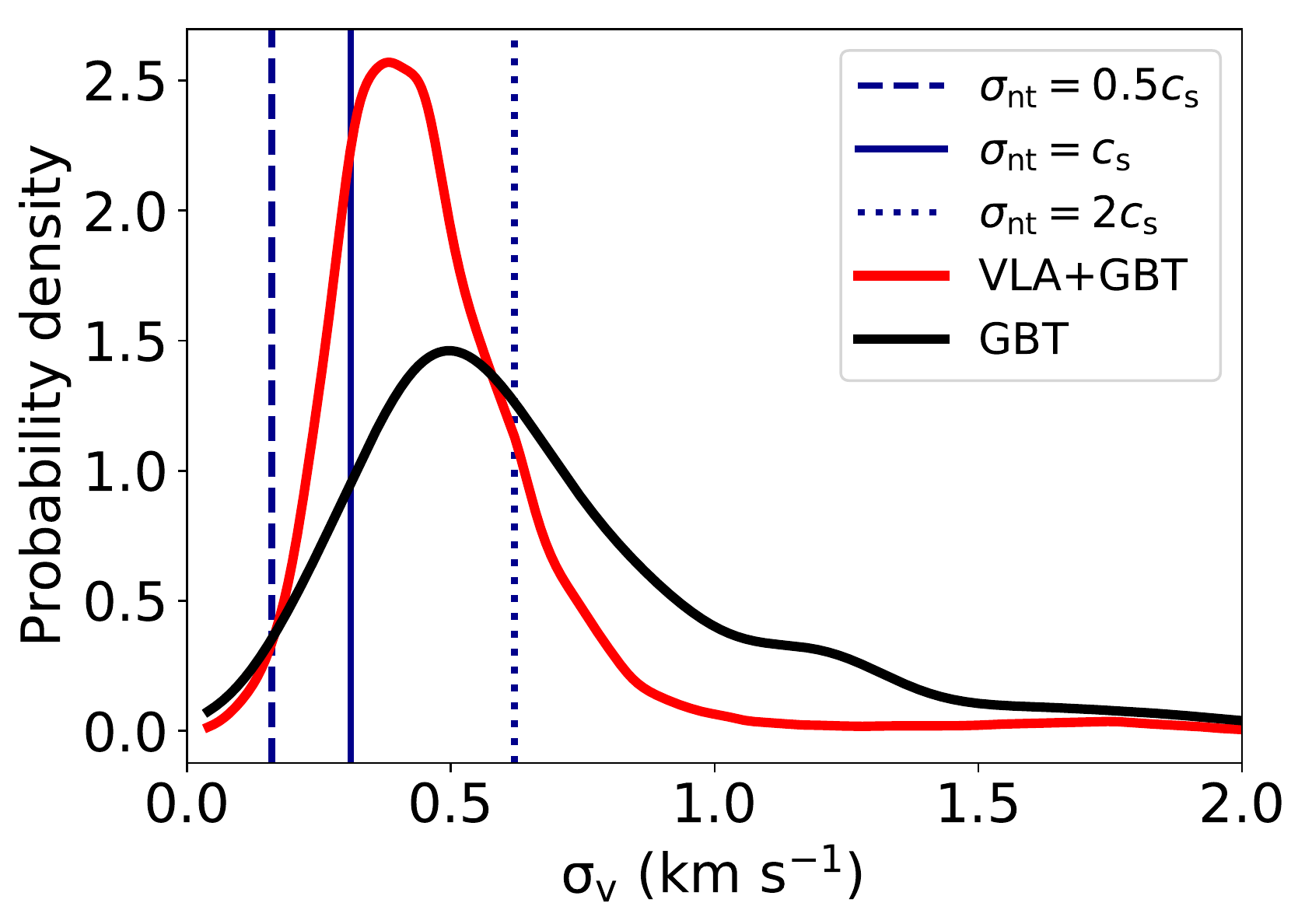}
\caption{Distribution of the corrected velocity dispersion (red curve) compared to the one obtained from GAS (black curve). The blue lines correspond to the expected non-thermal velocity dispersion for $\Tk=27.1~\K$ and $\sigma_{\mathrm{nt}}=0.5~c_{\mathrm{s}}$ (dashed), $\sigma_{\mathrm{nt}}=c_{\mathrm{s}}$ (solid) and $\sigma_{\mathrm{nt}}=2.0~c_{\mathrm{s}}$ (dotted) at $0.16~\kms$, $0.31~\kms$ and $0.62~\kms$, respectively. The sound speed of the gas $c_{\mathrm{s}}$ was calculated using a mean molecular weight of $\mu=2.33$, as described in detail in \S\,\ref{sec:kinematics}.}
  \label{fig:hist_dv}
\end{figure}

For comparison, the distribution function of the corrected velocity dispersion $\dv$ in our combined data is presented in Figure~\ref{fig:hist_dv}, together with that obtained from the GAS data over the same region. Immediately apparent is the shifted peak of the velocity dispersion toward narrower line widths for our higher-resolution data. 
The lower spatial resolution of the GAS data likely causes the blending of small-scale structures with narrow line widths, which not only broadens the total distribution of $\dv$ but shifts it toward larger line widths. 

The velocity dispersions $\dv$ of the $\NH$ gas derived from our data are defined as $\dv=\sqrt{\sigma^2_{\mathrm{t}}+\sigma^2_{\mathrm{nt}}}$, where $\sigma_{\mathrm{t}}$ and $\sigma_{\mathrm{nt}}$ correspond to the thermal and non-thermal velocity dispersions of the gas. Since $\sigma_{\mathrm{t}}$ is independent of the species under consideration, we assume that it is the same for the bulk of the gas with typical ISM conditions. The non-thermal component, however, does depend on the species via 

\begin{equation}
\sigma_{\mathrm{nt}}=\mathcal{M}\cdot c_{\mathrm{s}}=\mathcal{M}\cdot \sqrt{\frac{k_{\mathrm{B}}T_{\mathrm{K}}}{\mu m_{\mathrm{H}}}},
\end{equation}
where $\mathcal{M}$ is the Mach number, $c_{\mathrm{s}}$ the sound speed, $\mu = 2.33$ the mean molecular weight assumed for the ISM (i.e., molecular hydrogen and helium constitute most of the gas), $m_{\mathrm{H}}$ the mass of a hydrogen atom, $k_{\mathrm{B}}$ the Boltzmann constant and $T_{\mathrm{K}}$ the kinetic temperature of the $\NH$ gas. 

To study whether the gas in OMC1 is dominated by thermal or non-thermal motions, we assume a fixed $\Tk=27.1~\K$ (median kinetic temperature of OMC1 derived from the GAS data) for three different Mach numbers, corresponding to the subsonic ($\mathcal{M}=0.5$), sonic ($\mathcal{M}=1$) and supersonic ($\mathcal{M}=2$) regimes, for which we obtain $0.16~\kms$, $0.31~\kms$ and $0.62~\kms$, respectively.
These are plotted in Figure~\ref{fig:hist_dv} and one can clearly see that the $\dv$ distribution from GAS peaks in the supersonic regime.
In contrast, the combined VLA \& GBT data sample the non-thermal motions at smaller scales, finding that the velocity dispersion is only slightly supersonic at this higher angular resolution. It is therefore highly suggestive that observations with both higher angular and spectral resolutions than currently available would reveal even narrower line widths reaching subsonic turbulence.

\subsection{Integrated intensity}
\label{sec:mom0}

\begin{figure*}[]
\includegraphics[width=\linewidth]{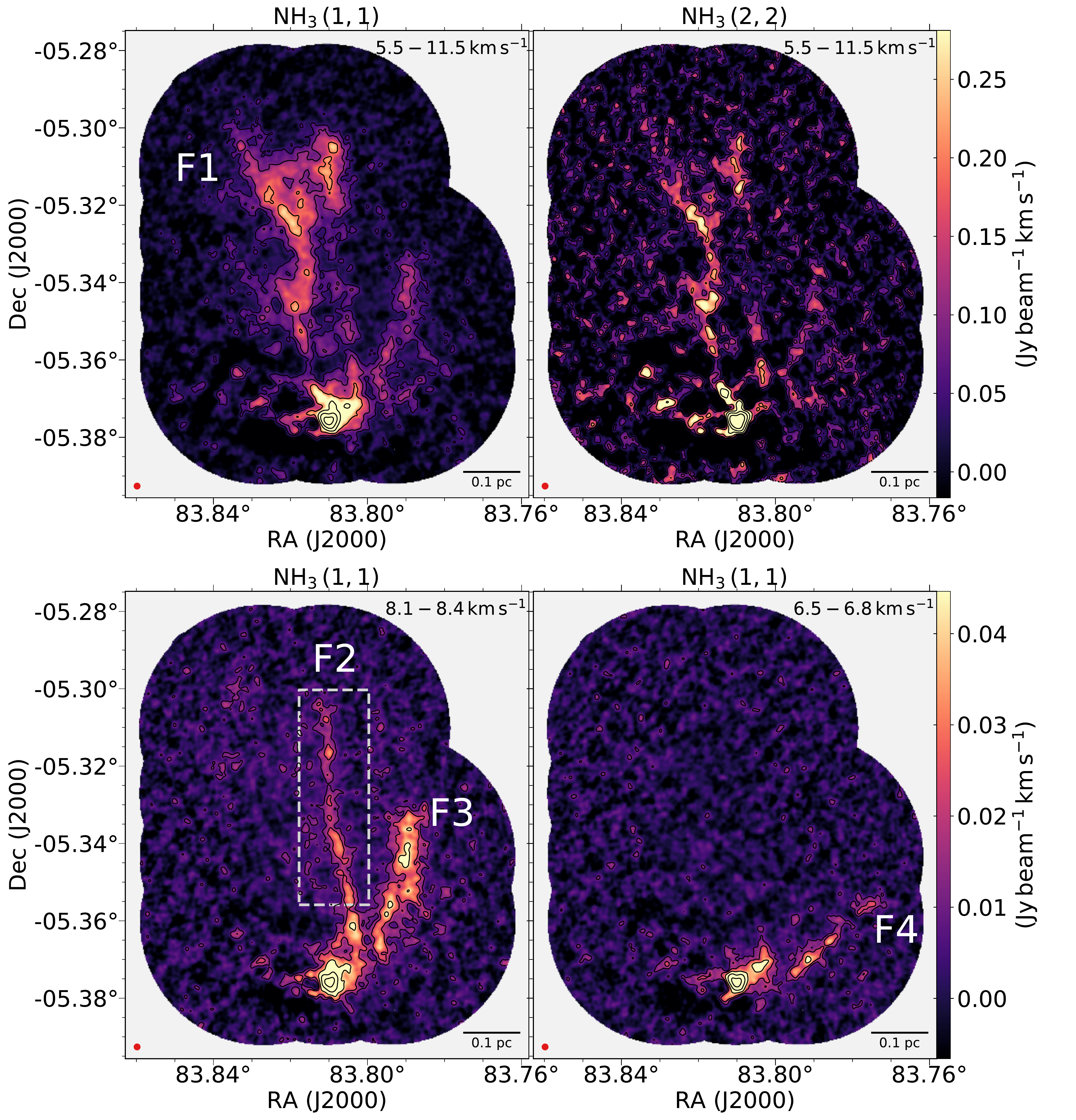}
\caption{\textit{Top:} $\NH\,(1,1)$ and $(2,2)$ integrated intensity maps of the combined data cube, calculated within $5.5$\,--\,$11.5~\kms$ to illustrate the overall gas distribution in OMC1. 
\textit{Bottom:} $\NH\,(1,1)$ integrated intensity maps for two different spectral ranges, to highlight the faint emission of the filaments F2, F3 (left, $8.1$\,--\,$8.4~\kms$) and F4 (right, $6.5$\,--\,$6.8~\kms$). The dashed white box surrounds F2, for which a more detailed analysis is presented in \S\,\ref{sec:mom0}. The beam size and spatial scale are given in the lower left and right corners of each panel, respectively. All contours are drawn at the $[3, 6, 12, ...]-\sigma_{\mathrm{rms}}$ levels of the corresponding map.}
  \label{fig:mom0_comp}
\end{figure*}

The combined $\NH\,(1,1)$ and $(2,2)$ integrated intensity maps of our data are shown in direct comparison in the upper panels of Figure~\ref{fig:mom0_comp}. The velocity range of $5.5$\,--\,$11.5~\kms$ used for calculating them includes the emission of the bright main component only. The $\NH$ distribution in our maps is in good agreement with previous studies of the same region \citep{WisemanHo1996_nature, WisemanHo1998}, but our observations have wider coverage of spatial scales, as discussed in \S\,\ref{sec:imgcomb}. Compared to the lower angular resolution GBT data alone, an increase in small-scale structures is readily seen and complex filamentary structures appear that extend in a finger-like arrangement from the bright Orion BN/KL region in the south. 
The $\NH\,(2,2)$ integrated intensity matches closely that of the $(1,1)$ line, although with lower SNR due to the higher temperatures needed to excite the $\NH\,(2,2)$ state and the resulting fainter emission. Therefore, the $\NH\,(2,2)$ emission is indicative of the presence of warmer gas in OMC1, especially along the most prominent filament labeled as F1 and Orion BN/KL in the south. Moreover, the finger-shaped filaments extending toward the northwest of OMC1 also show a substantial amount of bright emission in $\NH\,(2,2)$.

To highlight these faint filaments, we additionally show $\NH\,(1,1)$ integrated intensity maps in the lower two panels of Figure~\ref{fig:mom0_comp}, each covering only two specific channels, in which their emission is strongest. While the integrated intensity map between $5.5$\,--\,$11.5~\kms$ is ideal for tracing the overall, larger-scale gas content of the region, the emission of the brightest and most prominent filament F1 overwhelms the fainter structures in that region. F2 to F4 are only clearly visible in the narrow spectral ranges between $8.1$\,--\,$8.4~\kms$ and $6.5$\,--\,$6.8~\kms$ (both covering two channels), partly explaining why some of them were missed in previous studies at lower spectral resolution. To further illustrate the channel-specific appearance of F2, we additionally show a set of $\NH\,(1,1)$ channel maps, covering the velocity range $\sim7.4$\,--\,$9.6~\kms$ in Figure~\ref{fig:channel_map} in Appendix \ref{sec:app_channelmaps}.

\subsection{Filament profile}
\label{sec:F2_profile}

\begin{figure}[]
\centering
\includegraphics[angle=270, width=\linewidth]{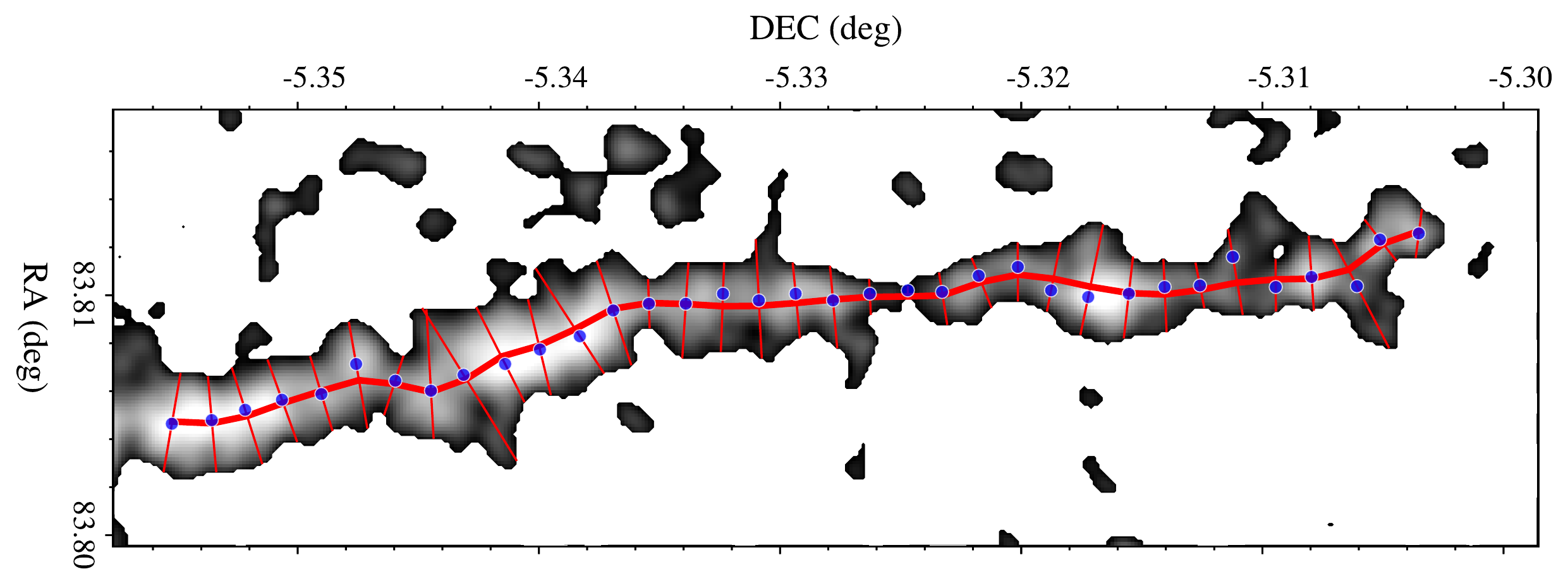}
\caption{Upper $25^{\mathrm{th}}$ percentile of F2's integrated intensity emission between $8.1$\,--\,$8.4~\kms$. The filament spine (thick, red line) was identified using the package \texttt{FilFinder} \citep{KochRosolowsky2015_filfinder}. The filament profile was then built using \texttt{RadFil} \citep[][]{Zucker+2017, Zucker+2018} by sampling radial cuts (short, red lines) every $0.01~\pc$ along the spine. The radial distance is then defined to be the projected distance from the peak emission at a given cut (blue dots).}
  \label{fig:F2_cuts_radfil}
\end{figure}

The narrow filament F2 that connects the southern and northern parts of OMC1 is framed by a white, dashed box in Figure~\ref{fig:mom0_comp} to show the region on which we will focus in the following analyses. Due to its faint emission in the full integrated intensity map, F2 was not identified in any previous observations of OMC1. However, its emission is strongest within the narrow velocity range of $8.1$\,--\,$8.4~\kms$, allowing the precise measurement of F2's width and length.

We determined the mean $\NH\,(1,1)$ integrated intensity profile of F2 using functionality from the python packages \href{https://github.com/e-koch/FilFinder}{\texttt{FilFinder}}\footnote{\href{https://github.com/e-koch/FilFinder}{https://github.com/e-koch/FilFinder}} \citep{KochRosolowsky2015_filfinder} and \href{https://github.com/catherinezucker/radfil}{\texttt{RadFil}}\footnote{\href{https://github.com/catherinezucker/radfil}{https://github.com/catherinezucker/radfil}} \citep[][]{Zucker+2017, Zucker+2018}. First, we defined a mask for F2 by computing the $75^{\mathrm{th}}$ percentile of intensity values shown inside the white box in Figure~\ref{fig:mom0_comp} (bottom left), and masking out all pixels with an intensity value below this level. 
Then, to find the ``spine'' of F2 we performed medial axis skeletonization of this mask using the \texttt{FilFinder} package, which reduces the filament mask to a one-pixel wide representation of the mask's topology. In Figure~\ref{fig:F2_cuts_radfil}, this spine (thick red line) is shown overlaid on the filament mask (background grayscale). Then, using the \texttt{RadFil} package, we determined the line tangent to the spine at $0.01~\pc$ intervals along the filament, and took the radial cut perpendicular to each tangent. These radial cuts are shown as thin red intersecting lines in Figure~\ref{fig:F2_cuts_radfil}. Since the spine is not weighted by integrated intensity, we shift each cut so that the peak is centered at $0~\pc$, where the peak is determined using all the points along each radial cut within the filament mask. The pixels with the peak intensity along each cut are shown as blue points in Figure~\ref{fig:F2_cuts_radfil}, and the radial profiles are built out from these points. Given the beam size of $6''$, this results in an ensemble of statistically independent cuts, where the radial distance is defined as the projected distance from the peak intensity at a given cut.

\begin{figure}[h]
\centering
\includegraphics[width=\linewidth]{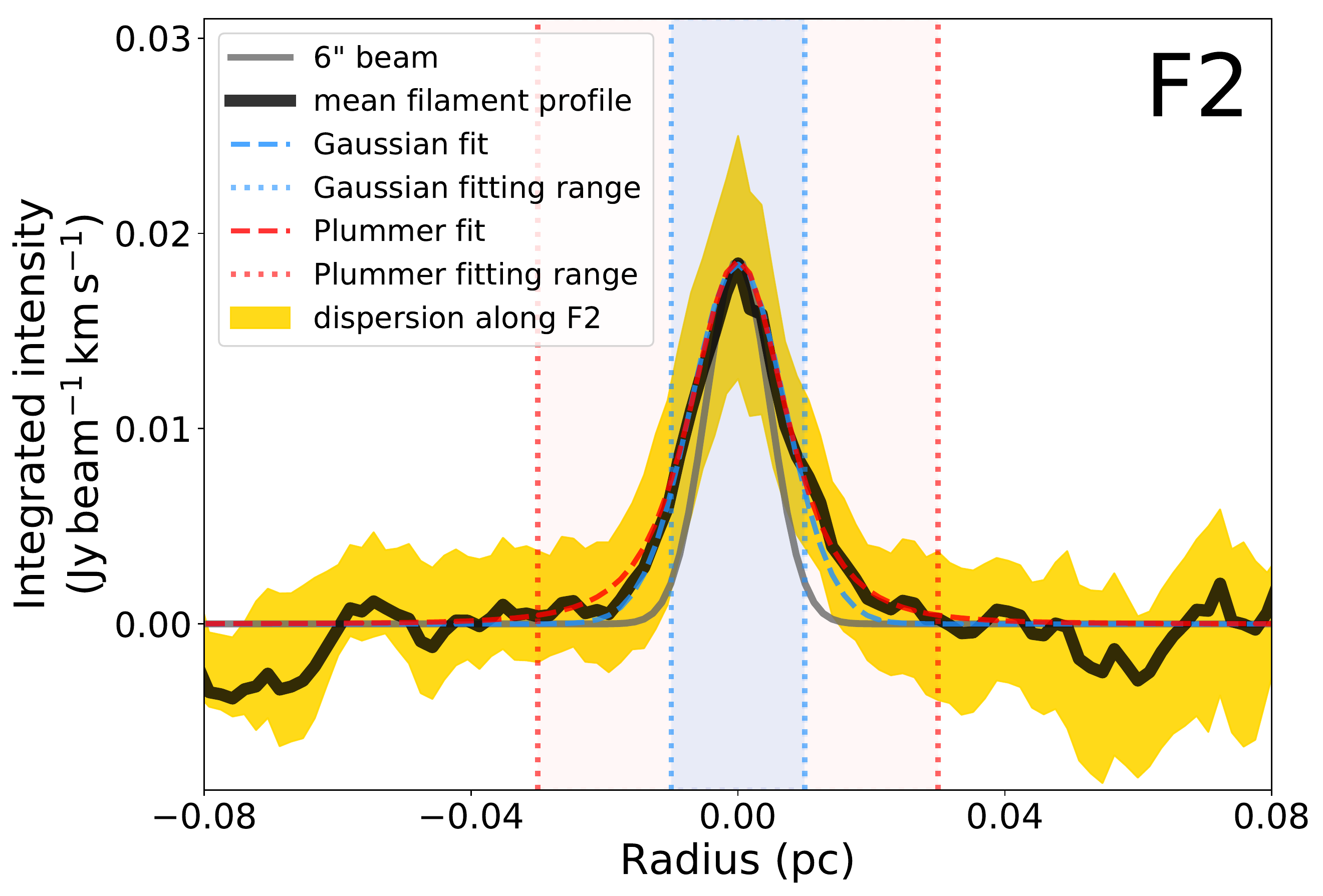}
\caption{Mean integrated intensity profile of the averaged filament profile of F2 as function of radial distance, calculated between $8.1$\,--\,$8.4~\kms$ (solid, black line) after the background emission has been subtracted. The yellow shade corresponds to the standard deviation of the cuts at each radial distance. The solid, gray line shows the beam response with a FWHM of $6''$. The dotted, blue line corresponds to a Gaussian fit with a deconvolved FWHM of $0.012~\pc$, and the red, dashed line to a Plummer fit with $p=5.1$ and $R_{\rm{flat}}=0.013~\pc$. The vertical, dotted lines show the fitting ranges for the Gaussian (blue) and Plummer (red) profile, respectively. The best fit parameters for both profiles are listed in Table~\ref{table:param}.}
\label{fig:F2_width}
\end{figure}

To fit the profiles, we used \texttt{RadFil}'s built-in background subtraction estimator, along with its Gaussian and Plummer fitting functions. While we performed all fitting on the ensemble of radial cuts we show the mean profile (solid black line) and the standard deviation of the cuts at each radial distance (yellow shaded area) in Figure~\ref{fig:F2_width}. Before performing the Gaussian and Plummer fitting, we first estimate the background, which is then subtracted off from each cut by fitting a first-order polynomial to all profiles at radial distances between $0.03~\pc$ and $0.05~\pc$. Then, we fit a Gaussian to the ensemble of cuts using all points within a radial distance of $0.01~\pc$. 
Our best-fit Gaussian, for which we list the best-fit parameters in Table~\ref{table:param}, is shown in blue in Figure~\ref{fig:F2_width} and corresponds to a deconvolved FWHM\footnote{The deconvolved FWHM is determined using the formula from \citet{Konyves+2015}: $\mathrm{FWHM_{deconv}=\sqrt{FWHM^2 - HPBW^2}}$, where HPBW is the half-power beamwidth in parsecs. The HPBW of our observations is $6''$ (equivalent to $0.011~\pc$ at the distance to Orion), and the solid gray line in Figure~\ref{fig:F2_width} shows the response of our $6''$ Gaussian beam.} of $0.012~\pc$ at a distance of $(388\pm5)~\pc$ to Orion \citep{Kounkel+2017}. We find the uncertainty on our FWHM value by individually
fitting each cut with a Gaussian, as opposed to fitting the entire ensemble of cuts. The uncertainty of $0.003~\pc$ we report in Table~\ref{table:param} is the standard deviation of all the deconvolved FWHM values we determine via the individual fits. With a measured length of $0.45~\pc$ and a deconvolved FWHM of $(0.012\pm0.003)~\pc$, this corresponds to an aspect ratio of approximately 37:1 for F2.

An alternative width can be obtained by assuming that F2 is a cylindrical filament with a Plummer-like density profile, an approach commonly used in the analysis of submillimeter observations, such as those from \textit{Herschel} \citep[e.g.,][]{Nutter+2008, Arzoumanian+2011, Palmeirim+2013, Smith+2014, Salji+2015}. Assuming that the integrated intensity, $\int T\mathrm{d}V(r)$, is a good tracer of the column density, we fit the following relation to the radial profile: 
\begin{equation}
\int T\mathrm{d}V(r) = \frac{A_0}{\left[ 1+\left(r/{R_{\rm{flat}}} \right)^2 \right]^{(p-1)/2}},
\end{equation}
where $r$ is the radial distance, $R_{\rm{flat}}$ the radius of the inner, flat region of the density profile, $A_0$ is the peak profile amplitude, and $p$ the power-law density exponent at large radii \citep{Cox+2016, Zucker+2017}. To cover the region at which the Plummer profile flattens, a larger fitting range of $\pm0.03~\pc$ for the entire ensemble of cuts was used, which is highlighted by the vertical red lines in Figure~\ref{fig:F2_width}.

\begin{deluxetable}{lcccc}
\tablecolumns{5}
\tablewidth{\columnwidth}
\tablecaption{Best fit parameters for the radial intensity profile of F2. \label{table:param}}
\tablehead{
\colhead{Type} & \colhead{Amplitude} & \colhead{Width\tablenotemark{a}} & \colhead{Exponent} & \colhead{$\mathrm{FWHM_{deconv}}$}\\
\colhead{} & \colhead{($10^{-2}\,\frac{\mathrm{Jy}}{\mathrm{beam}}\frac{\mathrm{km}}{\mathrm{s}}$)} & \colhead{($10^{-2}\,$pc)} & \colhead{} & \colhead{($10^{-2}\,$pc)}}
\startdata
Gaussian & $1.8\pm0.5$ & $0.7\pm0.2$ & \nodata & $1.2\pm0.3$\\
Plummer & $1.9\pm 0.6$ & $1.3\pm 0.5$ & $5.1\pm1.0$ & $2.0\pm0.8$\\
\enddata
\tablenotetext{a}{Width corresponds to the standard deviation ($\sigma$) in the Gaussian profile and to the flattening radius ($R_{\rm{flat}}$) in the Plummer profile. The deconvolved FWHM is then calculated using the relation $\mathrm{FWHM_{deconv}=\sqrt{FWHM^2 - HPBW^2}}$ from \citet{Konyves+2015}, where $\mathrm{FWHM}=2\sqrt{2\ln{2}}\cdot \mathrm{width}$ and HPBW is the half-power beamwidth in parsecs (i.e. $0.011~\pc$, for a beam size of $6''$).}
\end{deluxetable}

After subtracting off the background in the same manner as for the Gaussian fit, we obtain $p=5.1$ for the power-law exponent of the Plummer profile, which is larger than previously found for filaments studied in Cygnus or Taurus \citep[e.g.][]{Arzoumanian+2011, HacarTafalla2011, Palmeirim+2013}. The radius $R_{\rm{flat}}=0.013~\pc$ of the inner flat region of F2 is, however, an order of magnitude lower than typically seen in such studies. In fact, the width of F2 is more similar to the narrow filament found in the Barnard 5 region in Perseus, which was discovered in combined data at the same angular resolution of 6$\arcsec$ \citep{Pineda+2011,Pineda+2015}. 
We summarize the full set of best-fit parameters from both profiles in Table~\ref{table:param}.

\begin{figure*}[]
\centering
\includegraphics[width=\linewidth]{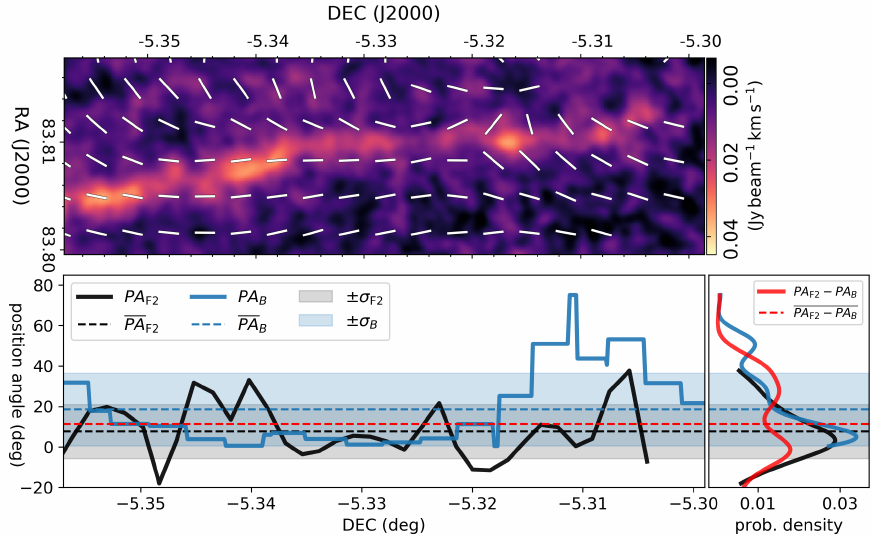}
\caption{\textit{Top:} Zoom-in into the white box in the integrated intensity map shown in Figure~\ref{fig:mom0_comp} with overlaid $B$-field half-vectors from \citet{Pattle+2017, Ward-Thompson+2017}. \textit{Lower left:} Position angles of F2 ($PA_{\rm{F2}}$, black curve) and the nearest $B$-field vector to F2 ($PA_{B}$, blue) as a function of declination. \textit{Lower right:} Distribution of $PA_{\rm{F2}}$ (black), $PA_{B}$ (blue) and their corresponding angle difference $PA_{\rm{F2}}-PA_{B}$ (red), obtained using a Gaussian kernel density estimate implemented in \texttt{SciPy} \citep{scipy}. The dashed lines and shades show the mean position angle and its standard deviation for the corresponding distribution.}
  \label{fig:PA}
\end{figure*}
To quantify the straightness of F2, we used \texttt{RadFil} to measure the variation of F2's curvature along its length. The position angle ($PA$) of F2 is then defined to be the upper angle between a horizontal (east-west) line placed on the filament spine and each cut.
The mean position angle of F2 is found to deviate by only $\sim8\degree\pm 13\degree$ from a perfectly straight filament extending north-south, where the uncertainty of $\pm 13\degree$ describes the standard deviation of F2's mean position angle. 

The finding of F2 is especially interesting considering recent observations of the dust polarization from the \textit{James Clerk Maxwell} Telescope (JCMT) BISTRO Survey \citep{Pattle+2017, Ward-Thompson+2017}. The authors show maps of the polarization half-vectors rotated by $90\degree$ to illustrate the orientation of the magnetic field lines in OMC1 (henceforth ``$B$-field half-vectors'', where ``half-vector'' refers to the $\pm180\degree$ ambiguity in magnetic field direction, as explained in detail by the authors). This map is overlaid onto SCUBA-2 $850~\rm{\mu m}$ observations of Orion~A at $14.1''$ resolution.
Due to the resolution of $6''$ achieved in our $\NH$ maps, i.e. better by a factor of $\sim2$, we can already infer from a visual inspection of both data sets that the magnetic field lines are oriented almost entirely parallel to F2. 
To investigate this connection further, we plot in the upper panel of Figure~\ref{fig:PA} the JCMT BISTRO $B$-field half-vectors onto the zoomed-in $\NH\,(1,1)$ integrated intensity map between $8.1$\,--\,$8.4~\kms$. The lower left panel of Figure~\ref{fig:PA} shows a direct comparison of the position angles measured at each cut along F2 and the nearest $B$-field half-vector as a function of declination. As already described, we find a mean position angle of $\sim8\degree\pm 13\degree$ for F2 and a slightly larger $PA$ of $\sim18\degree\pm18\degree$ for the $B$-field. In the lower right panel, we show the resulting distribution of position angles for both F2 and the $B$-field half-vectors, obtained using a Gaussian kernel density estimate\footnote{The kernel density estimate (KDE) is an estimator for the probability density function of a distribution, like, e.g., a histogram. While the latter depends both on the starting position and the bin width, the KDE solely depends on the selected bandwidth. Therefore, the KDE exhibits in general better convergence properties than the more common histogram and additionally estimates a smooth function \citep[e.g.,][]{BashtannykHyndman2001,Scott1992}.} (KDE) in \texttt{SciPy}. 
In order to compare both distributions directly, we determine the position angle difference of F2 and the $B$-field ($PA_{\rm{F2}}-PA_{B}$) along F2's spine and plot its KDE onto the distribution in the lower right panel of Figure~\ref{fig:PA}. The deviation between $PA_{\rm{F2}}$ and $PA_{B}$ is small and peaks near $0\degree$, reaching a mean value of $\sim11\degree$. This confirms that F2 and the $B$-field are aligned almost parallel to each other. However, higher deviations of $> 20\degree$ are found between declinations of $\delta\sim -5.32\degree$ and $-5.31\degree$ surrounding the higher-density condensation in F2.
We finally note that in order to do a proper comparison between two different sets of observations, one should ideally smooth those to the same angular resolution. In this case, however, smoothing the $\NH\,(1,1)$ integrated intensity map of the combined data to the JCMT beamsize of $14.1''$ makes it hardly possible to identify F2 in this map due to its low SNR. The subsequent  analysis of straightness or orientation compared to the $B$-field would then not be possible. Future observations with improved sensitivity should help to resolve this issue.

Finally, we estimate the total column density of F2 to lie between $10^{22.5-23}~\cm^{-2}$ \citep[i.e. $\sim 30$\,--\,$100~\mathrm{mag}$, see][]{Bohlin+1978,Pineda+2008} by using the derived $N(\NH)$ column densities from the fit and a $[\NH/\mathrm{H_2}]$ abundance of $10^{-8.5}$ to $10^{-8.0}$ \citep{FriesenPineda+2017}. This column density value is 2 to 3 orders of magnitude larger than those of \textit{Herschel} filaments studied in Cygnus or Taurus, and is comparable to what is typically observed from the inner parsec scale regions around high-mass star-forming regions \citep[e.g.,][]{Lin+2017}. Therefore, F2 is likely self-gravitating, given the low values of turbulence found, and could then fragment into smaller condensations to further allow for the star-formation in the region to proceed \citep[see][for a similar fragmentation in a low-mass star forming region]{Pineda+2015}. 

\section{Discussion}
\label{sec:discussion}

\subsection{F2 filament profile}
\label{sec:F2}

The results presented in this paper show that complex filamentary structures are present even on the smallest scales in OMC1. As already pointed out, this region has been the subject of numerous studies, but most of them either focused on the large-scale structure of Orion~A or the local environment of the Orion BN/KL region. 
Filament F2, however, has not been previously identified in any of these studies, even in the high-resolution maps by \citet{WisemanHo1996_nature, WisemanHo1998}. 
This finding is particularly interesting in the context of previous \textit{Herschel} observations of the Gould Belt star-forming regions \citep[e.g.,][]{Andre+2010, Arzoumanian+2011, Palmeirim+2013} that have found filaments to be ubiquitous structures in molecular clouds. Furthermore, the \textit{Herschel} filaments are found to exhibit a narrow distribution of widths, peaking at approximately $0.1~\pc$, often referred to as the ``characteristic width'' of filaments. \citet{Panopoulou+2017}, however, have recently argued that this width is not a universal attribute of filaments, but a result of the applied method of measuring filament widths. They find that this presumably ``characteristic'' width results from averaging the filament profiles along the spine. However, the filament width at each cut can vary from one end of the filament to the other (c.f. Figure~\ref{fig:F2_cuts_radfil}). Fitting an averaged filament profile instead of the entire ensemble of profiles may therefore artificially broaden the filament profile, resulting in the overestimation of its width. 
This is, however, not the case for F2, as fitting the averaged profile would only result in a slight increase of the filament width, namely from $(0.012\pm0.005)~\pc$ to $(0.013\pm0.005)~\pc$.
Further, \citet{Smith+2014} emphasize that the range used for fitting the filament column density profile may also strongly influence the resulting width. Therefore, it should be chosen small enough to avoid the possible contribution of overlapping filaments \citep{Federrath2016}. 
In the case of F2, adopting a larger fitting radius of $0.02~\pc$ and $0.03~\pc$ results in deconvolved FWHMs of $(0.015\pm0.003)~\pc$, i.e. only up to a factor of $\sim0.2$ larger than the width determined using a fitting radius of $0.01~\pc$. 
The reported filament width of $0.012~\pc$ is therefore only a lower limit.

We find qualitatively similar behavior for the Plummer fit, which is less affected when changing the fitting range. For fitting radii of $0.02~\pc$ or $0.04~\pc$, we find $p=5.0$, $R_{\mathrm{flat}}=0.013~\pc$ and $p=5.4$, $R_{\mathrm{flat}}=0.014~\pc$, respectively.  

Additionally, the observation of narrow filaments with widths less than $0.1~\pc$, like F2 in OMC1, in Barnard 5 in Perseus \citep{Pineda+2011, Pineda+2015}, or in Serpens South \citep{Fernandez-Lopez+2014}, seem to confirm that there is no universality in `characteristic' filament widths \citep[see also][who investigate the effect of magnetic fields on filament widths depending on their orientation]{SeifriedWalch2015}. 
We stress, however, that the $\NH$ hyperfine transitions specifically trace denser and cooler gas than \textit{Herschel} dust emission. Therefore, narrower filament widths may be naturally expected to be measured using $\NH$ as a tracer. For example, \citet{FriesenPineda+2017} find a filament in NGC~1333 that is a factor of 1.4 narrower in $\NH$ than in \textit{Herschel} dust emission. Nevertheless, the resolution of $\sim 18''$ with \textit{Herschel} and $32''$ with the GBT are significantly larger than the $6''$ achieved in the ammonia maps presented here. The lack of filaments with widths below $0.1~\pc$ in star-forming regions may therefore partly stem from the scarcity of high-resolution observations of dense molecular gas tracers such as $\NH$ or $\rm{N_2H^+}$.
As dust continuum observations do not routinely achieve the resolution needed to resolve such narrow filaments, more interferometric observations of high-density gas tracers are clearly required. 

F2 is well fitted by a Plummer-like density profile with $p=5.1$, which is larger than previously found for filaments studied in Cygnus or Taurus where typically flatter power-law exponents ($1.5 < p < 2.5$) are found \citep{Arzoumanian+2011}.
Such filaments are in disagreement with the model of an isothermal filament in hydrostatic equilibrium, where steeper profiles with a power-law exponent $p=4$ are expected \citep{Ostriker1964}. Potential explanations for these flatter filament profiles may be the support by magnetic fields \citep[e.g.,][]{FiegePudritz2000a, FiegePudritz2000b}, their formation in turbulent clouds \citep{Smith+2014} or the inclusion of external pressure \citep[e.g.,][]{FischeraMartin2012} into the \citet{Ostriker1964} solution \citep{Federrath2016, Smith+2014, Lada+1999}. However, F2 is found to have an even larger power-law exponent of $p=5.1$ but a flattening radius of $R_{\rm{flat}}=0.013~\pc$, which is about ten times smaller than those of typical \textit{Herschel} filaments. This value is more similar to the widths of filaments identified in $\NH$ by \citet{Pineda+2011, Pineda+2015}, suggesting that the filaments revealed in these studies and in ours appear to have different properties from the ones previously studied. 

We finally note that F2 is apparently aligned with one of the CO streamers resulting from the explosive outflow seen in Orion BN/KL \citep[c.f. 'Finger \#2' in Figures 7 and 8 in][]{Bally+2017}. It is highly suggestive that the massive explosion that occurred in Orion BN/KL about 500 years ago has not only shaped the dense gas in that region but also influenced the magnetic field. In contrast, \citet{Hacar+2017} suggest that OMC1 is still gravitationally collapsing and argue that the origin of the molecular fingers seen in OMC1 might be the result of the velocity gradient along each of these structures and reflect the kinematic structure of the gas on large scales.

\subsection{Orientation of F2 in comparison to the magnetic field in OMC1}
\label{sec:orientation}

We have shown in \S\,\ref{sec:mom0} that F2 has a mean position angle of $PA=8\degree\pm 13\degree$ with generally little variation along its length. The polarization half vectors from the JCMT BISTRO survey show that the magnetic field along F2 is oriented almost entirely parallel to this filament.
The largest deviations between the position angles of F2 and the magnetic field half-vectors can be found in the northern section of F2, i.e. at declinations $> -5.32\degree$. 
As can be inferred from the integrated intensity map in Figure \ref{fig:mom0_comp}, the bright filament F1 outshines the faint emission of F2 in that region. 
Therefore we expect the magnetic field to trace the denser filament F1 rather than F2 in this part. 

Previous \textit{Herschel} studies confirmed numerical calculations \citep[e.g.,][]{Nagai+1998} which have shown that a parallel orientation between filaments and the local magnetic field is to be expected for filaments having low column densities \citep[e.g.,][]{Peretto+2012, Palmeirim+2013, Andre+2014, Busquet+2013, Cox+2016, Panopoulou+2016}. 
In fact, a variety of orientations have previously been observed and the alignment of filaments and apparent field direction appear to vary between regions and for different gas densities \citep[c.f.][for a review]{Pillai2017}. For example, there is observational evidence that low-column density filaments, or ``striations'', are aligned parallel to the magnetic field lines in Taurus, as they are more susceptible to the magnetic influence than higher column density structures \citep{Palmeirim+2013}. These striations may therefore promote mass accretion onto larger filaments, in which stars are finally formed. Such studies, however, were mostly performed using lower-resolution dust column density maps obtained with \textit{Herschel} \citep[e.g.,][]{Li+2013, Palmeirim+2013, Soler+2017} or polarization data tracing relatively low $A_{\mathrm{V}}$ environments 
\citep[e.g.][]{Busquet+2013, Fissel+2016, Planck+2016, Soler+2016, Jow+2017}.
Nevertheless, as discussed in \S\,\ref{sec:F2_profile}, the filament F2 studied in this work has molecular hydrogen column densities roughly 2 to 3 orders of magnitude larger than the \textit{Herschel} filaments studied in Cygnus or Taurus and may therefore possibly be self-gravitating.
At face value, these results are at odds with the widely accepted paradigm of magnetically-regulated star formation in filaments, in which a perpendicular orientation of dense filaments with respect to the magnetic field is expected \citep[e.g.,][]{Peretto+2012, Palmeirim+2013, Andre+2014, Busquet+2013, Cox+2016, Panopoulou+2016}.

The relative orientation between the polarization vectors and filaments is the most basic and simplest to compute and recent work has suggested several statistics to interpret the observations and compare robustly with numerical simulations \citep[e.g.,][]{Soler+2013, SolerHennebelle2017}. 
To date, a variety of simulations have explored the properties of magnetized filaments, and it is clear that the nature of the alignment between filaments and magnetic fields is inseparable from the role magnetic fields play in the formation of filaments \citep[e.g.,][]{SeifriedWalch2015}. For example, if the gas is weakly magnetized and filaments are produced by shocks, the magnetic field may be disordered, which in turn reduces the effective polarization and impacts the apparent direction of the field \citep{Hull+2017, King+2017, Seifried+2018}. If the magnetic field is weak in F2, the relative orientation of F2 and the dust polarization we find might not indicate the real orientation of the magnetic field. While this is possible, we note that the direction of the polarization vectors change in the vicinity of the main over-density in the filament (at $\delta \sim -5.32\degree$, see Figure \ref{fig:PA}). Combined with the fact that the magnetic field strength of $B = 6.6\pm4.7~\mathrm{mG}$ determined by \citet[][]{Pattle+2017} in the entire OMC1 is strong and in agreement with different studies of OMC1 \citep[e.g.,][]{Hildebrand+2009}, this suggests that the magnetic field along F2 is indeed reflecting the gas distribution and unlikely to be completely dynamically negligible.  

Synthetic observations further show that the viewing angle with respect to the filament and field direction also has a large effect on the apparent relative orientation. For example, in colliding flow simulations where the field is initially inclined at $20\degree$ with respect to the direction of flow, \citet{King+2017} found the field to be enhanced in the dense post-shock region and its bulk orientation to be changed. An observer looking along the direction of the flow would perceive the field to be primarily in the plane-of-sky. However, an observer with an edge-on view of the shock, would perceive a strong component along the line-of-sight. In fact, the synthetic polarization vectors of the edge-on view of the \citet{King+2017} super-Alfvenic colliding flow model looks very similar to our F2 observations. \citet{Chen+2017} examined the impact of velocity perturbations in colliding flow simulations on the resulting field morphology. They found that perturbations varying spatially along the field can produce striations, in which the field is ``rolled up'', thereby producing a field aligned with the filament rather than perpendicular to it. This situation also resembles our observations of F2. Without 3D spatial information, however, it is difficult to infer information about the gas morphology for F2. Its sub-sonic turbulent line width suggests it is unlikely that the gas is very extended along the line of sight.

Finally, it is possible that the orientation of the field and filament changes with time such that the field appears more parallel at earlier times and more perpendicular at later times \citep{Chen+2016}.
\citet{ChenOstriker2015} estimate the time for collapse to proceed to densities of 10$^7$ cm$^{-3}$ starting from a mean density of 10$^3$ cm$^{-3}$ as $0.4$\,--\,$0.8~\Myr$. If this is the natural progression, F2 may be fairly young and the polarization vectors will gradually anti-align with the field as the filament gains mass. A higher velocity in the core compared to the filament could also point to this scenario \citep{Chen+2016},  but this is not currently apparent from the $\NH$ data. However, the fact that the higher density condensation within the filament at $\delta \sim -5.32\degree$ shows some misalignment is suggestive.

In conclusion, observations of magnetic fields in dense, i.e. high-$A_{\mathrm{V}}$ filaments in (sub-)millimeter bands at high angular resolution have unveiled the coexistence of parallel and perpendicular configurations between $B$-fields and filaments. Therefore, filaments are very likely to be formed in more than one way and a uniform formation theory cannot reproduce the observed variation of angle differences. Observations at high-angular resolution using dense gas tracers such as $\NH$ will help to identify similar filaments like F2 and broaden our understanding of magnetically regulated filament formation in star-forming regions.
\section{Summary}
\label{sec:summary}

The results presented in this paper reveal the density structure of dense gas in OMC1 in Orion~A in unprecedented detail. The combination of single-dish and interferometric observations with both high spatial and spectral resolutions resulted in a beamsize of 6$\arcsec$, allowing to resolve structures as small as 0.01$~\pc$ at the distance of 388$~\pc$ to OMC1. 
These high-quality observations of the $\NH~(1,1)$ and $(2,2)$ hyperfine transitions allow detailed analyses of the gas distribution of the densest and coolest parts of OMC1.
The main results are summarized as follows:

\begin{itemize}
\item A narrow filament named F2 with a deconvolved FWHM of $0.012~\pc$ and an aspect ratio of $\sim$37:1 has been revealed. Owing to the improved images produced by combining existing VLA data with recent single-dish GBT data, F2 could be identified for the first time.

\item F2 is found to be of high-column density, corresponding to visual extinctions $A_{\mathrm{V}}\sim30$\,--\,$100~\mathrm{mag}$. This is several orders of magnitude larger than the filaments and striations previously studied with \textit{Herschel} and suggests that F2 is possibly self-gravitating.

\item F2 is a relatively straight filament extending north-south with a mean position angle of $8\degree\pm13\degree$. Compared to the mean orientation of $18\degree\pm18\degree$ for the magnetic field along F2, a mean deviation of $\sim 11\degree$ was measured, showing that both vector fields are aligned effectively parallel to each other. This situation is at odds with the currently widely accepted paradigm of magnetically-supported filament formation, in which a perpendicular orientation of a dense filament with respect to the magnetic field is expected.

\item The mean velocity dispersion measured in the combined VLA \& GBT data is about $0.2~\kms$ smaller than previously determined in lower-resolution observations. This difference suggests that higher angular and spectral resolution observations will identify even lower velocity dispersions, that may even reach subsonic turbulence.

\end{itemize}

\acknowledgments
We thank the anonymous referee for a thorough and constructive review and Markus Michael Rau for his help on improving this paper. 
K.M. and B.E. acknowledge the support of the DFG priority program SPP-1992 ``Exploring the Diversity of Extrasolar Planets'' (DFG PR 569/13-1, ER 685/7-1) \& the DFG Research Unit ``Transition Disks'' (FOR 2634/1, ER 685/8-1) and the Munich Institute for Astro- and Particle Physics (MIAPP) of the DFG Cluster of Excellence `Origin and Structure of the Universe'. 
J.E.P. and P.C. acknowledge the financial support of the European Research Council (ERC; project PALs 320620).
The JCMT data used in this paper were obtained under project codes M16AL004 and M15BEC02. K.P. acknowledges support from the Science and Technology Facilities Council (STFC) under grant number ST/M000877/1.
This paper makes use of archival data of the \textit{Robert C. Byrd} Green Bank Telescope (GBT) and the \textit{Karl G. Jansky} Very Large Array (VLA) operated by the National Radio Astronomy Observatory (NRAO). The National Radio Astronomy Observatory is a facility of the National Science Foundation operated under cooperative agreement by Associated Universities, Inc.
The \textit{James Clerk Maxwell} Telescope is operated by the East Asian Observatory on behalf of The National Astronomical Observatory of Japan; Academia Sinica Institute of Astronomy and Astrophysics; the Korea Astronomy and Space Science Institute; the Operation, Maintenance and Upgrading Fund for Astronomical Telescopes and Facility Instruments, budgeted from the Ministry of Finance (MOF) of China and administrated by the Chinese Academy of Sciences (CAS), as well as the National Key R\&D Program of China (No. 2017YFA0402700). Additional funding support is provided by the Science and Technology Facilities Council of the United Kingdom and participating universities in the United Kingdom and Canada.

\facilities{\textit{Karl G. Jansky} Very Large Array (VLA), \textit{Robert C. Byrd} Green Bank Telescope (GBT), \textit{James Clerk Maxwell} Telescope (JCMT)}

\software{\texttt{APLpy} \citep{RobitailleBressert2012}, \texttt{Astro\-py} \citep{astropy}, \texttt{Matplotlib} \citep{Hunter2007}, \textsc{Casa} \citep{CASA}, \texttt{FilFinder} \citep{KochRosolowsky2015_filfinder}, \textsc{Miriad} \citep{Miriad_software}, \texttt{NumPy} \citep{Numpy}, \texttt{PySpecKit} \citep{pyspeckit}, \texttt{RadFil} \citep[][]{Zucker+2018}, 
\texttt{RadioBeam} \citep{pythontools_radiobeam}, \texttt{scikit-image} \citep{scikit-image}, \texttt{SciPy} \citep{scipy}, \texttt{SpectralCube} \citep{spectral-cube} }

\bibliographystyle{yahapj}
\bibliography{references}

\appendix

\section{Channel maps of NH$_3$ (1,1)}
\label{sec:app_channelmaps}

\begin{figure*}[h!]
\centering
\includegraphics[width=\linewidth]{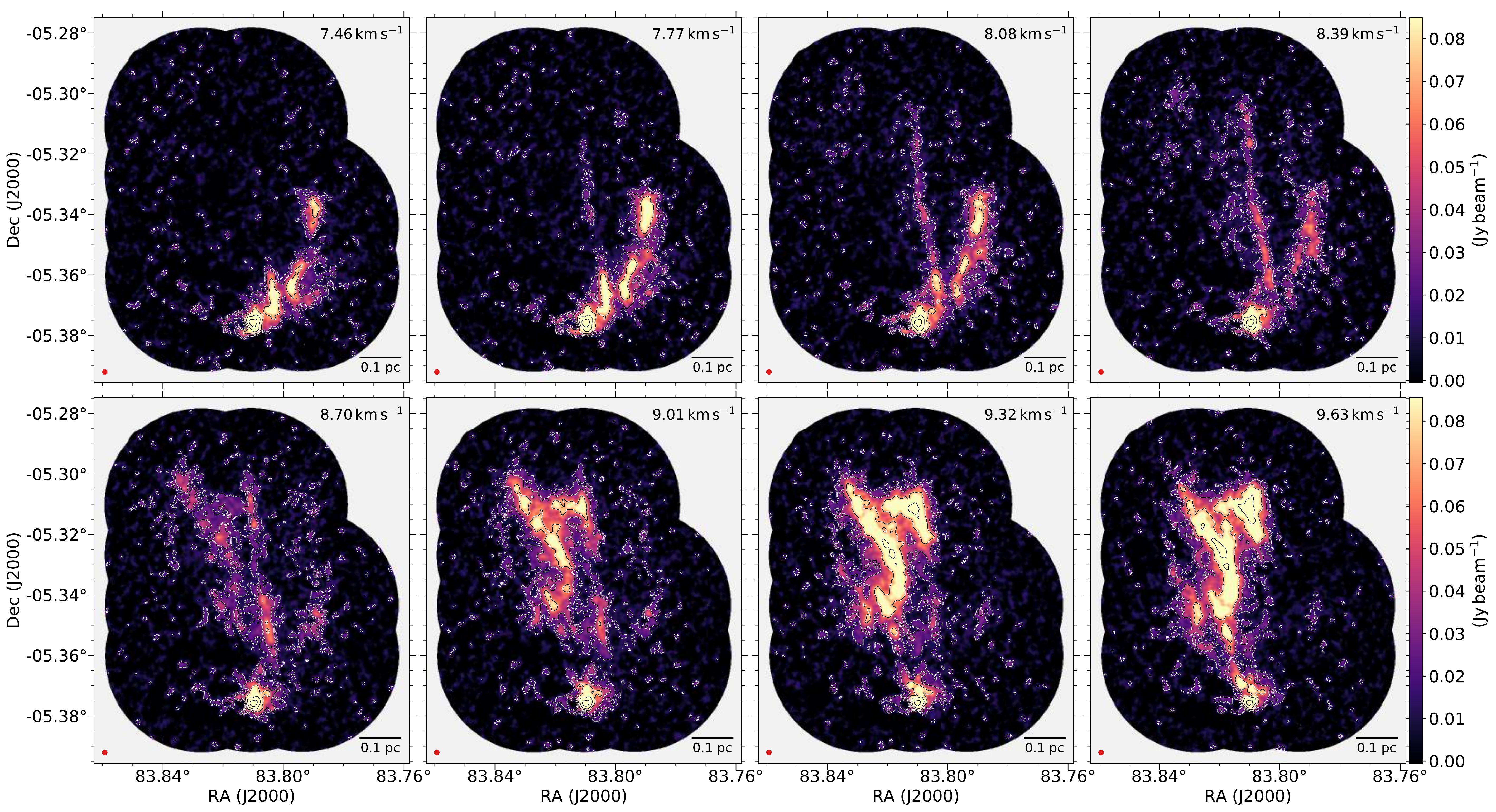}
\caption{Channel maps of the $\NH\,(1,1)$ transition with contours at the $[3, 6, 12, ...]-\sigma_{\mathrm{rms}}$ levels for each channel. The central velocity of each channel and the image scale are shown in the upper and lower right corners of each panel, respectively. }
    \label{fig:channel_map}
\end{figure*}

Figure \ref{fig:channel_map} shows channel maps of the $\NH\,(1,1)$ transition centered on $7.46~\kms$ to $9.63~\kms$ with a channel width of $0.3~\kms$ to highlight the channel-specific appearance of filament F2, which is discussed in detail in \S~\ref{sec:mom0}. It is mostly outshined by the bright emission of F1, and only visible within a narrow spectral range. 

\newpage

\section{Temperatures and $\NH$ column density}
\label{sec:temperatures}

Other parameters that result from the spectral line fit are the kinetic temperature $\Tk$, the excitation temperature $\Tex$, and the logarithmic ammonia column density $\log_{10}(N(\NH))$. 
At gas temperatures $\gtrsim 30~\K$, temperature estimates obtained from the $\NH\,(1,1)$ and $(2,2)$ only become inaccurate as higher-order $\NH\,(J=K)$ states get significantly populated.
Additionally, to determine the kinetic temperature of the gas, reliable fits of both the $\NH\,(1,1)$ and $(2,2)$ line profiles are required. The $\NH\,(2,2)$ emission is, however, very weak and only detected in regions with strong $\NH\,(1,1)$ emission. Furthermore, the rms-errors of the $\NH\,(2,2)$ transition are about four times higher than for $\NH\,(1,1)$, resulting in a large amount of pixels where the kinetic temperature could not be estimated accurately. As a consequence of the conservative flagging criteria pointed out in \S\,\ref{sec:methods}, only a small amount of pixels with reliable fits of the temperatures remain. This low yield especially applies for $\Tex$ and $\log_{10}(N(\NH))$, as they both are masked out if $\Tk$ is poorly constrained. 
Therefore we exclude these results from the scientific discussion, as more sensitive observations of $\NH\,(1,1)$, $(2,2)$ and higher-order transitions are clearly required to study the temperature structure in OMC1 in more detail. 

\begin{figure*}[h!]
\centering
\includegraphics[width=\linewidth]{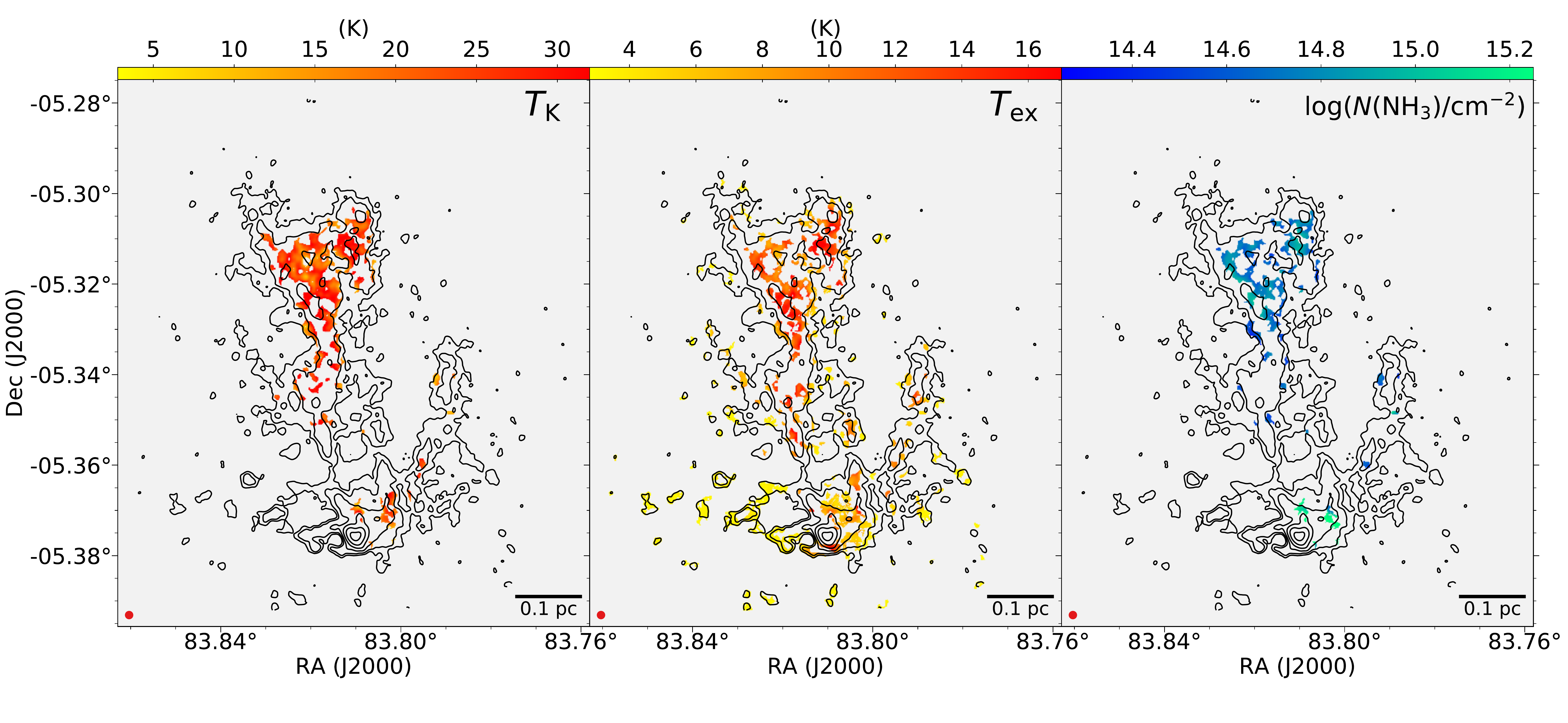}
\caption{Results of the spectral line fit, after pixels with uncertain fits were removed using the flagging conditions as listed in Table \ref{tab:flagging}. \textit{Left:} kinetic temperature $\Tk$. \textit{Center:} excitation temperature $\Tex$. \textit{Right:} logarithmic $\NH$ column density $\log(N(\NH)/\mathrm{cm^{-2}})$. The contours are drawn at the $[3, 6, 12,...]-\sigma_{\mathrm{rms}}$ levels from the $\NH\,(1,1)$ integrated intensity map of the final data cube.}
    \label{fig:fit_temp}
\end{figure*}

Figure \ref{fig:fit_temp} shows the resulting fits of the kinetic temperature $\Tk$ (left), the excitation temperature $\Tex$ (center) and the logarithmic $\NH$ column density $\log(N(\NH)/\mathrm{cm^{-2}})$ (right) after pixels with uncertain fits, as described in \S~\ref{sec:methods}, have been masked out.

\end{document}